%% file: Revision_noblue_main.tex
\documentclass[lettersize,journal]{IEEEtran}
\usepackage[table]{xcolor}
\usepackage{soul,framed} 
\usepackage{comment}
\colorlet{shadecolor}{yellow}
\usepackage[pdftex]{graphicx}
\graphicspath{{../pdf/}{../jpeg/}}
\DeclareGraphicsExtensions{.pdf,.jpeg,.png}

\usepackage[cmex10]{amsmath}
\usepackage{array}
\usepackage{mdwmath}
\usepackage{mdwtab}
\usepackage{eqparbox}
\usepackage{url}
\usepackage{graphicx}
\usepackage{amsmath}
\usepackage{amsfonts}
\usepackage{algorithm}
\usepackage{algpseudocode}
\usepackage{multirow}
\usepackage{multicol}
\usepackage{footnote}
\usepackage{bm}
\usepackage{makecell}
\usepackage{pdflscape}
\usepackage{afterpage}
\usepackage{bbding}
\usepackage{lscape}
\usepackage{booktabs}
\usepackage{tikz}
\usepackage{tabularx}
\usepackage{mathrsfs}
\usepackage[figuresright]{rotating}
\usepackage{subfigure}
\usepackage{amssymb}  
 \usepackage{threeparttable}
 \usepackage{caption}

\usepackage{pgfplots}
\pgfplotsset{compat=1.12}
    \pgfplotsset{
        layers/my layer set/.define layer set={
            background,
            main,
            foreground
        }{ },
        set layers=my layer set,
    }

\usepackage[square,sort,comma,numbers]{natbib} 

\usepackage[pagebackref=false,breaklinks=false,linkcolor=red,anchorcolor=blue, citecolor=green,colorlinks,bookmarks=true]{hyperref}
\usepackage{amsfonts}

\usepackage{pifont}

\begin{document}

\title{Fifty Years of SAR Automatic Target Recognition: The Road Forward}
 
\author{\IEEEauthorblockN{Jie Zhou},
Yongxiang Liu,
Li Liu,
Weijie Li,
Taoli Yang,
Bowen Peng,
Yafei Song,
Gangyao Kuang,
Xiang Li
\thanks{Jie Zhou, Yongxiang Liu, Li Liu, Weijie Li, Taoli Yang, Bowen Peng, Yafei Song, Gangyao Kuang and Xiang Li are with the College of Electronic Science and Technology, National University of Defense Technology (NUDT), Changsha, China. Taoli Yang is with the School of Resources and Environment, University of Electronic Science and Technology of China, Chengdu, China. Corresponding authors: Xiang Li (lixiang01@vip.sina.com), Yongxiang Liu (lyx\_bible@sina.com) and Li Liu (liuli\_nudt@nudt.edu.cn).}
\thanks{This work was supported by National Natural Science Foundation of China (NSFC) under Grant 62376283 and 62531026, Innovation Research Foundation of National University of Defense Technology, and the Fundamental and Interdisciplinary Disciplines Breakthrough Plan of the Ministry of Education of China (JYB2025XDXM110). Emails: Jie Zhou (zhoujie\_@nudt.edu.cn), Weijie Li (lwj2150508321@sina.com), Taoli Yang (yangtl@uestc.edu.cn) Bowen Peng (pbow16@nudt.edu.cn), Yafei Song (syf\_nudt@163.com) and Gangyao Kuang (kuangyeats@hotmail.com).}}

\markboth{In Preparation for IEEE GRSM}
{Zhou \MakeLowercase{\textit{et al.}}: }

\IEEEtitleabstractindextext{%

\begin{abstract}

Synthetic Aperture Radar (SAR)  imaging is capable of observing objects in nearly all weather and illumination conditions, and has become an indispensable means of information acquisition for analysis and recognition of objects and scenes. SAR Automatic Target Recognition (SAR ATR) has been one of the most fundamental and challenging problems in remote sensing image analysis. Nowadays, the artificial intelligence (AI) technology, represented by large models and AI agents, has transformed the research paradigm, profoundly influenced various research fields, and continues to evolve at an unprecedented pace. However, the huge potential of AI for SAR image analysis remains locked. To unlock the potential of AI in SAR image understanding, the research community should rethink how to enable bidirectional empowerment between AI and SAR image understanding and strive to achieve substantial breakthroughs at critical bottlenecks. Given this period of remarkable evolution, this paper offers the first comprehensive review of SAR ATR, tracing its development and milestones over the past five decades and providing the research community with a clear roadmap. This survey includes approximately 260 research contributions, covering critical aspects of SAR ATR: pivotal challenges, important datasets, the merits and limitations of representative methods, evaluation metrics, and state-of-the-art performance. Finally,  we finish the survey by identifying promising directions for future research. Looking ahead, we call for significant attention on three fundamental pillars: the curation of high-quality large-scale datasets, the design of fair and comprehensive evaluation benchmarks, and the fostering of safe open-source ecosystems.


\end{abstract}
}

\maketitle
\IEEEdisplaynontitleabstractindextext
\IEEEpeerreviewmaketitle

\input{Revision_noblue_section/introduction}

\input{Revision_noblue_section/problemOfSARATR}

\input{Revision_noblue_section/HistoryOfSARATR}

\input{Revision_noblue_section/EvolutionOfDetection}

\input{Revision_noblue_section/EvolutionOfSARClassification}
\input{Revision_noblue_section/RecentAdvances}

\input{Revision_noblue_section/Futurework}

\footnotesize
\bibliographystyle{IEEEtran}
\bibliography{IEEEabrv,myreference}

\input{Revision_noblue_section/bio}


\end{document}

%% file: Revision_noblue_section/introduction.tex

\section{Introduction}

\begin{figure*}[htb]
\centering
	\includegraphics[width=0.98\textwidth]{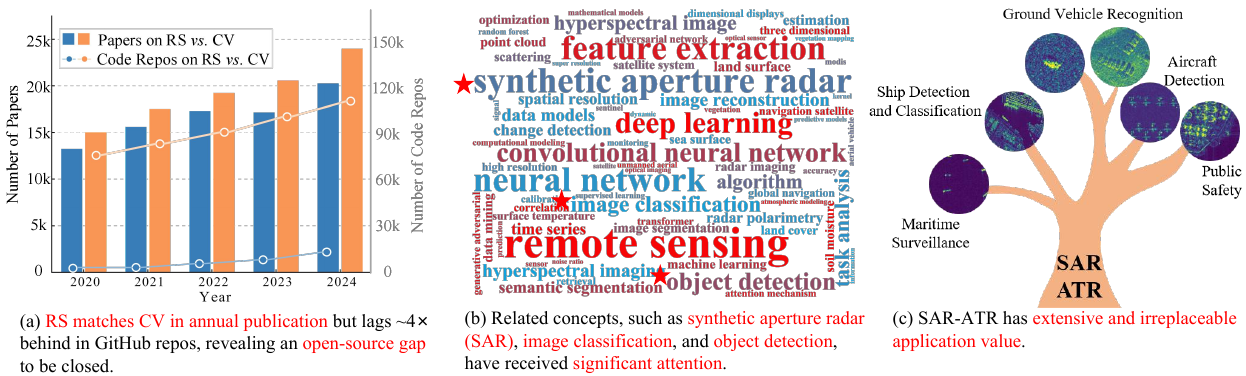}
	\caption{{\textbf{Importance of SAR ATR.} (a) From 2020 to 2024, the annual number of published papers in the fields of remote sensing (RS) and computer vision (CV) has reached a similar level (the gap is less than 10\%). However, the number of public RS-related code repositories on GitHub remains relatively limited, accounting for only approximately one-fourth of that in the CV domain. This discrepancy highlights significant untapped potential for advancing open-source ecosystem development within the remote sensing community. (b) Most frequent keywords in remote sensing-related papers from 2020 to 2024. The size of each word is proportional to the frequency, highlighting that concepts such as synthetic aperture radar (SAR), image classification, and object detection have garnered substantial attention. (c) SAR ATR has important application value in object-level target recognition tasks, including maritime surveillance, ship detection and classification, ground vehicle recognition, aircraft detection, and situational awareness in public safety and defense. As an important direction in target-level SAR image interpretation, SAR ATR has attracted sustained attention from both academia and industry.} (All statistics on the number of papers are from the WOS Core Collection Database.)}
 \label{wordcloud}
\end{figure*}

\subsection{Background}

Synthetic Aperture Radar Automatic Target Recognition (SAR ATR) aims to automatically detect and accurately classify targets of interest from SAR imagery, which are a unique type of all-time and all-weather ground observation data \cite{SAR2024SVA-CS-MPI,bodnar2025foundation}. In this review, we adopt SAR ATR in target-level sense, rather than broader SAR applications such as geophysical parameter retrieval or deformation monitoring. Since the launch of the first space-borne SAR satellite, Seasat-A, in 1978, this field has witnessed nearly five decades of development. The technological paradigms have undergone a profound evolution, from model-driven approaches based on physical scattering to optimization-driven methods based on statistical learning, and now to the current data- and knowledge co-driven stage \cite{SAR2000PAMIDet,SAR2012PAMISeg}. As an important component of SAR image interpretation, SAR ATR has consistently attracted substantial attention from both academia and industry (Fig. \ref{wordcloud}), with enduring significance from four interconnected aspects.

First, SAR ATR has important application value in object-level monitoring and recognition tasks. In maritime surveillance, it supports ship detection and classification for traffic management, fishery supervision, and navigation safety. In land and airborne surveillance, it enables vehicle and aircraft detection and recognition under complex weather and illumination conditions. It also plays an important role in public safety and national defense for situational awareness \cite{dong2022chronic}. Second, as a core technology for extracting target-level semantics from SAR imagery, advances in SAR ATR can also support broader progress in SAR scene understanding \cite{SAR2007PAMIRadar}. Third, the coherent imaging mechanism, speckle, geometric distortion, and complex scattering behavior of SAR imagery make visual interpretation and robust target recognition inherently challenging. The gap between microwave scattering characteristics and optical appearance is one of the key motivations for developing dedicated automated interpretation methods \cite{wu2025skysense}. Finally, SAR ATR lies at the intersection of multiple disciplines, attracting researchers from artificial intelligence (AI) \cite{fontanesi2025AI4Satellite,weng2024will}, pattern recognition \cite{jain2000statistical}, electromagnetics \cite{elachi1982spaceborne}, and signal processing \cite{reigber2012very,gomez2015multimodal}, and thus provides a valuable platform for cross-disciplinary innovation.

The field currently stands at a pivotal juncture, driven by the confluence of massive remote sensing data and rapid breakthroughs in large models, foundation models, and intelligent agents. This paper provides the first comprehensive review of SAR ATR, tracing five decades of progress and milestones to offer a clear roadmap for researchers.



\subsection{Comparison with Previous Surveys}

\begin{figure*}[t]
\centering
\includegraphics[width=0.99\textwidth]{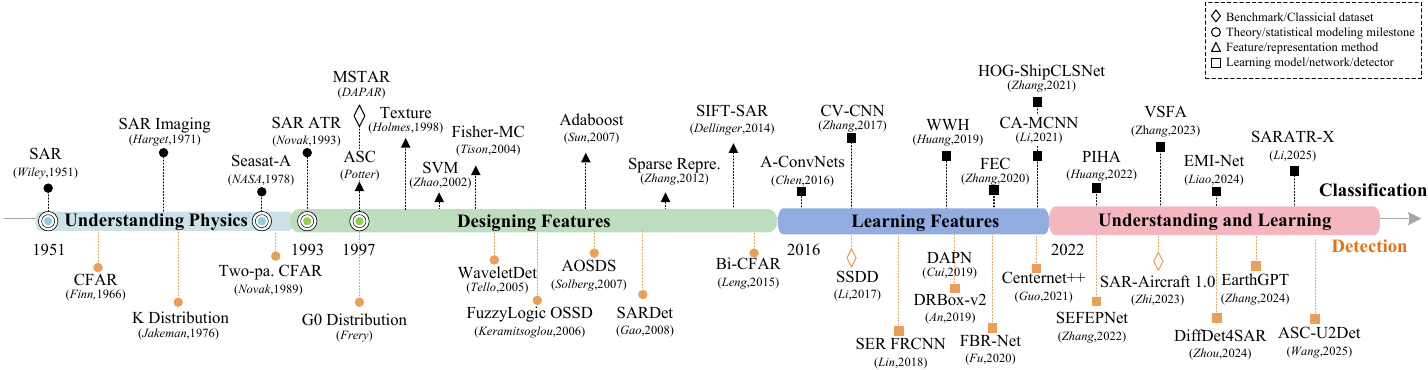}
\caption{Representative milestones in the evolution of SAR automatic target recognition (SAR ATR), organized around its two core tasks: classification (upper branch) and detection (lower branch). The timeline is not intended to be exhaustive; rather, the listed entries were selected based on their historical influence and paradigm representativeness. Because the figure aims to summarize field-shaping milestones rather than methods at a single technical level, each task branch intentionally includes benchmark datasets, theoretical/statistical modeling advances, feature representation methods, classifiers or detectors, and deep network architectures. The four developmental stages are organized according to the dominant target representation paradigm: understanding physics, designing features, learning features, and understanding-and-learning features. (\textbf{Classification:} SAR \cite{wiley1985synthetic}, SAR Imaging \cite{harger1971synthetic}, SAR ATR \cite{novak1993performance}, MSTAR \cite{keydel1996mstar}, ASC \cite{potter1997attributed}, Texture \cite{holmes1998textural}, SVM \cite{zhao2002SVMATR}, Fisher-MC \cite{tison2004newMK}, Adaboost \cite{sun2007adaboost}, Sparse Representation \cite{zhang2012multiJSR}, SIFT-SAR \cite{dellinger2014sarSIFT}, A-ConvNets \cite{chen2016target}, CV-CNN \cite{zhang2017complexCVCNN}, WWH \cite{huang2019WWH}, FEC \cite{zhang2020fec}, HOG-ShipCLSNet \cite{zhang2021hogShip}, CA-MCNN \cite{li2021CAMCNN}, PIHA \cite{huang2022piha}, VSFA \cite{zhang2023vsfa}, EMI-Net \cite{liao2024eminet}, SARATR-X \cite{li2025saratrx}. \textbf{Detection:} CFAR \cite{finn1966adaptiveCACFAR}, K Distribution \cite{jakeman1976KDistribution}, Two pa.CFAR \cite{novak1989studies}, Go Distribution \cite{frery1997model}, WaveletDet \cite{tello2005WaveletDet}, FuzzyLogic OSSD \cite{keramitsoglou2006autoOil}, AOSDS \cite{solberg2007oil}, SARDet \cite{2008gaoguidetectionsurvey,2009gaoguidiscriminationsurvey}, Bi-CFAR \cite{leng2015bilateralCFAR}, SSDD \cite{li2017shipssdd}, SER FRCNN \cite{lin2018SERFasterRCNN}, DAPN \cite{cui2019dapn}, DBBox-v2 \cite{an2019drboxV2}, FBR-Net \cite{Fu2020FERNet}, Centernet++ \cite{guo2021centernet++}, SEFEPNet \cite{zhang2022sefepnet}, SAR-AIRCraft 1.0 \cite{zhirui2023sarcraft}, DiffDet4SAR \cite{zhou2024diffdet4sar}, EarthGPT \cite{zhang2024earthgpt}, ASC-U2Det \cite{wang2025ASCU2Det}.)}
\label{History of SAR ATR}
\end{figure*}

Over the past few decades,the SAR ATR community has witnessed a tectonic shift from classical physics-based scattering models to modern deep learning paradigms. While numerous research milestones have significantly advanced the field (as illustrated in Fig. \ref{History of SAR ATR}), the existing literature remains largely fragmented. Although several reviews on SAR ATR exist, they tend to be either target-specific or technically narrow. Early retrospectives heavily favored physics-driven feature engineering \cite{2008gaoguidetectionsurvey,el2013targetSurvey}, whereas more recent summaries have pivoted strictly to task-specific deep-learning methodologies \cite{yasir2023shipSurvey,ru2023intelligentSurvey,2024OSGRSM}. To date, the community lacks a comprehensive review that traces this entire intellectual arc and integrates lessons from both eras. Critically, prior works have not explicitly addressed how the object-level semantic interpretation bridges physical imaging, nor have they revealed how classical physical insights are inherited, refined, and reformulated within contemporary deep learning architectures.

To bridge this critical gap and clarify the unique contributions of our manuscript, we summarize related works in TABLE \ref{table:survey_SUMMARY} and highlight the novelty of this survey across the following five distinguishing aspects:

\textbf{(i) Comprehensiveness:} Unlike existing fragmentary literature, this is the first review to systematically chronicle the complete fifty-year evolution of SAR ATR. It spans from the foundational statistical methods of the 1970s to the contemporary physics-integrated foundation models, distilling approximately 250 pivotal research contributions.

\textbf{(ii) Inheritance:} We uniquely reveal the intellectual continuity underlying paradigm shifts. Specifically, we analyze how traditional concepts, such as electromagnetic scattering center models and CFAR statistical detection, have not been discarded but actively assimilated and reimagined within modern deep learning frameworks.

\textbf{(iii) Systematicness:} We present a systematic and critical taxonomy of the principal challenges in SAR ATR. By explicitly distinguishing algorithmic limitations that have been substantially mitigated from those that persist as open research problems, we establish a rigorous framework for assessing and comparing the current state of the art.

\textbf{(iv) Openness:} To facilitate research reproducibility and support rapid prototyping, we construct a comprehensive, community-driven repository aggregating existing literature, open-source datasets, and codebases, accessible at: \href{https://github.com/JoyeZLearning/SAR-ATR-From-Beginning-to-Present}{https://github.com/JoyeZLearning/SAR-ATR-From-Beginning-to-Present}. We note that this repository is intended as a curated resource collection rather than a complete benchmark suite or a unified reference-implementation platform.


\textbf{(v) Future-oriented:} At a moment when large foundation models and AI agents are transforming research paradigms, we move beyond mere historic summarization. We chart a clear roadmap for the future, synthesizing key emerging directions including the curation of high-quality large-scale datasets, the design of fair and comprehensive evaluation benchmarks, and the fostering of safe open-source ecosystems.

\subsection{Scope and Organization}
Given the exponential growth of literature generated over the past half-century (as illustrated in Fig. \ref{wordcloud}(a)), achieving exhaustive coverage within a single manuscript is practically unattainable. To maintain depth, focus, and analytical rigor, we explicitly delineate the boundaries of this review along the following three dimensions:

\textbf{(i) Literature source:} peer-reviewed papers from high-impact top journals and conferences related to remote sensing, as well as works that have historically initiated paradigm shifts or introduced pioneering physical insights.

\textbf{(ii) Temporal span:} a continuous historical trajectory spanning nearly five decades from 1978, marked by the launch of Seasat-A, the first Earth-orbiting spaceborne SAR satellite, to the contemporary era of foundation models.

\textbf{(iii) Task scope:} object-level target detection and classification within single-channel, static SAR amplitude images, which constitutes the most fundamental and widely deployed domain format.

While advanced sub-fields such as moving target indication (MTI), Video SAR recognition, and fully polarimetric SAR (PolSAR) interpretation represent critical frontiers, their distinct underlying physical mechanisms warrant independent specialized reviews, and they are thus positioned as future expansions rather than the core subject of this review.

\begin{figure}[htbp!]

\centering
\includegraphics[width=0.4\textwidth]{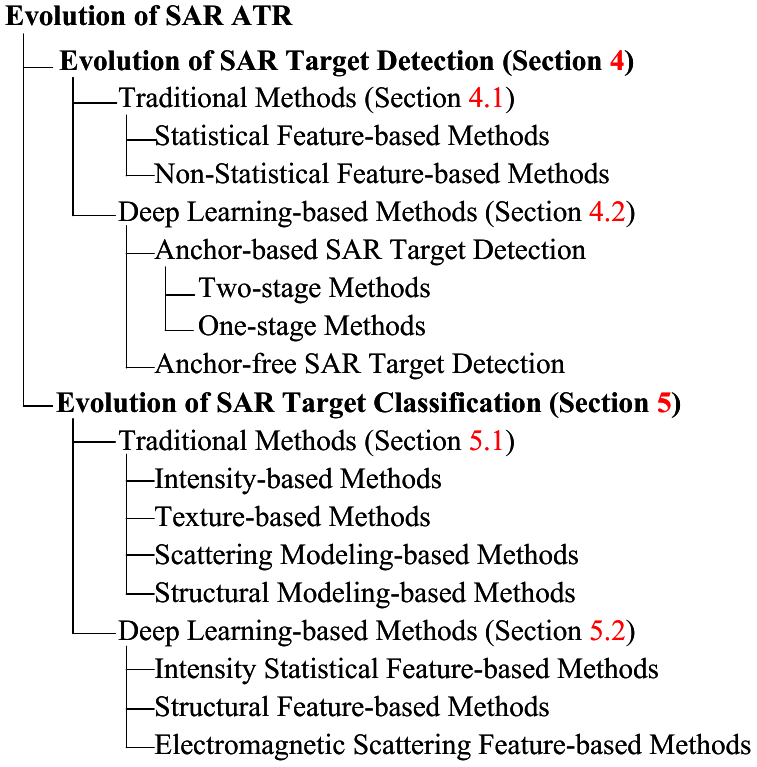}
\caption{The taxonomy of representative methods in SAR ATR.}
\label{overall framework}
 \vspace{-0.5cm}
\end{figure}

The remainder of this paper is structured as follows. Problem definition, core challenges, and datasets of SAR ATR are summarized in Section \ref{sec:problem}. In Section \ref{sec:history}, we review the history of SAR ATR. Section \ref{sec:detection} and \ref{sec:class} provide a comprehensive survey of the evolution of SAR target detection and classification. A taxonomy of SAR target detection and classification methods is illustrated in Fig. \ref{overall framework}. Section \ref{sec:recent advances} covers recent advances of SAR ATR. In Section \ref{sec:future}, we conclude the paper and discuss the possible promising future research directions.

\begin{table*}[!ht]
\centering
\caption{Summary of representative surveys in SAR ATR. \textbf{T} and \textbf{D} in the \textbf{Scope} column denote \textbf{Traditional} and \textbf{Deep learning-based} methods covered by each survey, respectively. The last two columns summarize the main focus of each survey and its corresponding scope characteristics.}
\label{table:survey_SUMMARY}
\resizebox{1\linewidth}{!}{
\footnotesize
\begin{tabular}{
>{\centering\arraybackslash}m{0.5cm}   
>{\centering\arraybackslash}m{0.6cm}   
>{\centering\arraybackslash}m{1.8cm}   
>{\centering\arraybackslash}m{0.7cm}   
m{6.3cm}  
m{6.3cm}  
}
\hline\hline
\textbf{Ref} & \textbf{Year} & \textbf{Topic} & \raisebox{-4pt}{\textbf{Scope}} & \textbf{Contributions} & \textbf{Scope Characteristics} \\
\hline
\cite{keydel1996mstar}
& 1996
& vehicle features
& T
& Emphasizes model-driven ATR and provides a detailed discussion of MSTAR EOCs and SAR target features.
& Primarily focuses on vehicle targets and feature analysis. \\
\hline
\cite{Crisp2004TheSI}
& 2004
& ship detection
& T
& Discusses the imaging mechanism and theoretical basis of SAR ship detection, together with implementation details and applications in real SAR images.
& Primarily centers on ship detection and related imaging mechanisms. \\
\hline
\cite{2008gaoguidetectionsurvey}
& 2008
& detection
& T
& Reviews traditional SAR target detection methods developed over the previous two decades based on contrast, image features, and complex-domain information.
& Focuses on detection methods. \\
\hline
\cite{2009gaoguidiscriminationsurvey}
& 2009
& discrimination
& T
& Reviews traditional target discrimination methods from the perspectives of feature extraction, prior knowledge, scattering characteristics, and angle-dependent variations between targets and clutter.
& Emphasizes discrimination-oriented analysis, with detection and classification discussed separately. \\
\hline
\cite{el2013targetSurvey}
& 2013
& CFAR detection
& T
& Categorizes SAR detection approaches into single-feature-based, multi-feature-based, and expert-system-oriented methods.
& Focuses specifically on CFAR and related detection methods. \\
\hline
\cite{el2016automaticSurvey}
& 2016
& holistic SAR ATR system perspective
& T
& Discusses SAR ATR from an end-to-end system perspective and proposes a two-fold benchmarking scheme for evaluating and motivating system design.
& Emphasizes holistic system evaluation, mainly using MSTAR-like benchmark settings rather than complex urban clutter or multi-target scenes. \\
\hline
\cite{dong2018reviewSARATRAE}
& 2018
& vehicle classification
& T
& Reviews SAR target classification from the perspective of autoencoders and their variants.
& Focuses mainly on vehicle classification and autoencoder-based methods. \\
\hline
\cite{lan2020SARsurvey}
& 2020
& detection
& T+D
& Surveys single-channel SAR target detection and identification methods in complex scenes over the past decade.
& Emphasizes detection and identification in complex scenes. \\
\hline
\cite{kechagias2021ATRSurvey}
& 2021
& classification on MSTAR
& T+D
& Summarizes SAR target classification methods on MSTAR from the perspectives of reflectance attributes and transformations.
& Focuses primarily on MSTAR-based classification. \\
\hline
\cite{li2022deepATRSurvey}
& 2022
& ship detection
& T+D
& Reviews 177 articles on SAR ship detection, covering deep learning frameworks and deployment-related considerations.
& Concentrates on ship detection. \\
\hline
\cite{yasir2023shipSurvey}
& 2023
& ship detection
& D
& Summarizes 81 deep learning-based SAR ship detection papers from 2016 to 2022, with emphasis on network architectures.
& Focuses mainly on deep network design for ship detection. \\
\hline
\cite{ru2023intelligentSurvey}
& 2023
& aircraft detection and classification
& D
& Delivers a comprehensive survey covering target characteristics, key challenges, algorithmic evolution, datasets, performance metrics, and future trends.
& Concentrates on aircraft-related tasks. \\
\hline
\cite{li2023RScomprehensive}
& 2023
& detection and classification
& D
& Reviews 197 papers from perspectives including small-sample learning, class imbalance, real-time processing, polarimetric SAR, and complex SAR.
& Provides broad deep-learning-oriented coverage. \\
\hline
\cite{lang2025recentATR}
& 2025
& detection and classification
& D
& Reviews 171 articles according to datasets and the detection/classification tasks of different target types.
& Focuses on recent deep learning studies. \\
\hline
\cite{slesinski2025review}
& 2025
& dual perspective for detection and classification
& T+D
& Reviews detection and classification methods from both traditional and deep learning perspectives, and emphasizes practical applications under real-time, lightweight, and on-device constraints.
& Provides a dual-perspective review with emphasis on practical deployment. \\
\hline
\cite{qiao2025ship}
& 2025
& ship interpretation
& D
& Reviews 418 papers published from 2022 to 2025, categorizes methods into optical-transferred and SAR-physics-customized branches, and sorts emerging tasks including wake detection and 3D refocusing.
& Covers the full spectrum of maritime SAR tasks with latest frontier works. \\
\hline
\cite{Muhammad2026JstarShip}
& 2026
& ship classification
& D
& Systematically analyzes 74 selected papers, establishes a four-dimensional taxonomy targeting classification, unifies OA and F1 for quantitative cross-model comparison, and summarizes core bottlenecks like data scarcity and long-tailed imbalance.
& Focus on single cropped SAR chip ship classification, covering airborne/spaceborne single/dual polarized SAR data and post-2010 papers with OA/F1 metrics. \\
\hline
\raisebox{-4pt}{\textbf{Ours}} 
& \textbf{2026}
& fifty-year evolution of SAR ATR
& \textbf{T+D}
& \textbf{Provides a broad review of fifty years of SAR ATR development, covering historical evolution, datasets, methods, challenges, and future directions from both traditional and deep learning perspectives.}
& \textbf{Focuses on a comprehensive historical and methodological overview, rather than exhaustive implementation-level detail or uniform experimental benchmarking of every method family.} \\
\hline\hline
\end{tabular}}
\end{table*}

%% file: Revision_noblue_section/problemOfSARATR.tex
\section{Problem of SAR ATR}
\label{sec:problem}

\subsection{Definition}

SAR ATR system was first proposed by Lincoln Laboratory in 1993 \cite{dudgeon1993overviewATR,novak1993performance,novak1994radarNovak}, with its classical architecture consisting of three progressively advanced stages: pre-screening, discrimination, and classification. In the prescreening stage, the system performs rapid processing on large-scene SAR images to eliminate background regions that obviously do not contain targets and outputs a number of Regions of Interest (ROIs) that may contain targets. The discrimination stage then conducts more refined analysis on these candidate regions to distinguish real targets from false alarms caused by natural clutter. In the classification stage, the system extracts discriminative features (e.g., scattering center distribution, contour moments) from the regions confirmed to be targets so as to realize the determination of specific categories, models and even identities (for example, distinguishing Boeing 737 from A330 aircraft).

\begin{figure*}[htb]

\centering
\includegraphics[width=0.93\textwidth]{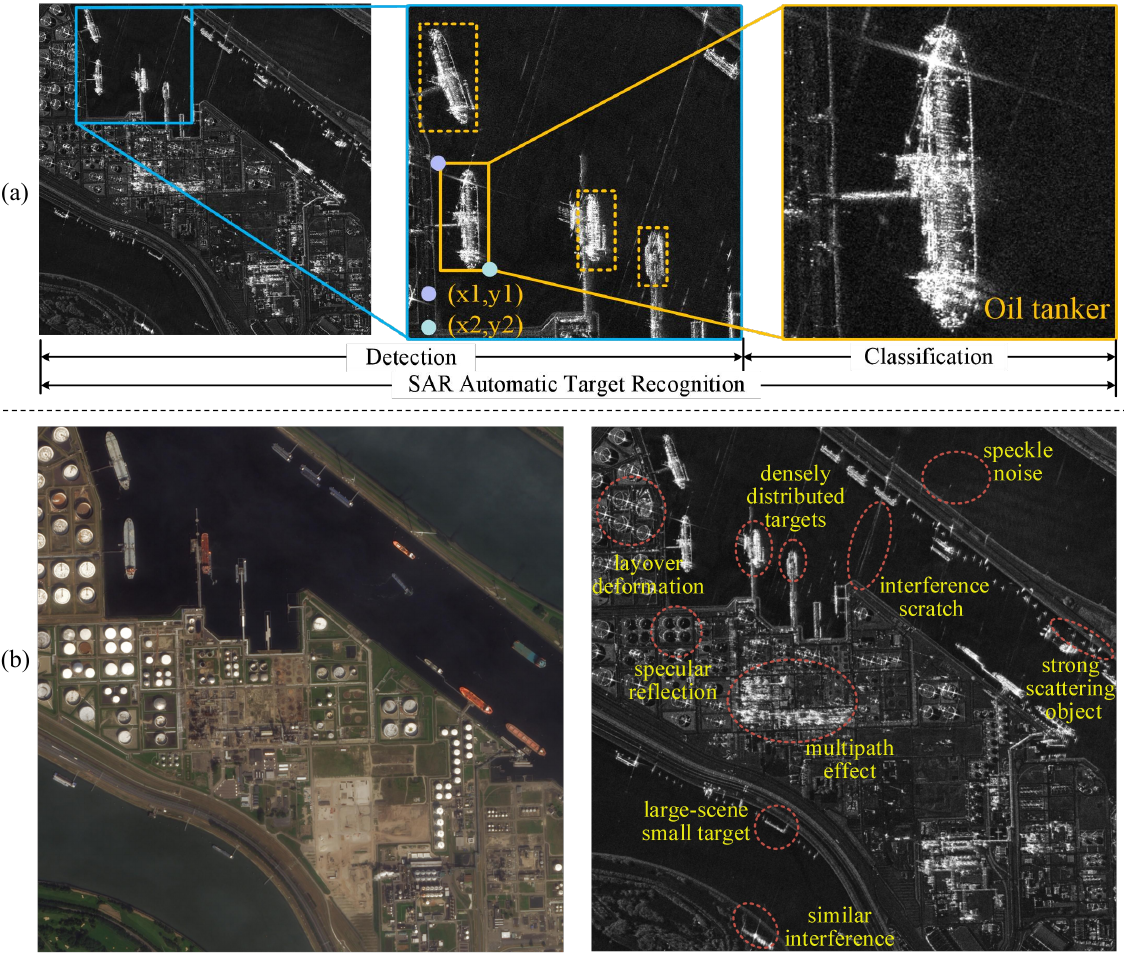}

\caption{
(a) Definition of SAR ATR. It encompasses two key stages: \textbf{detection}, which locates potential target regions within a large-scale SAR image, and \textbf{classification}, which classifies the specific category (exemplified by the oil tanker ship) of the detected target. (b) Difference between optical and SAR images, and some challenging instances during SAR target recognition.}
\label{definition and challenges of SAR ATRv2}

\end{figure*}

Over the more than two decades of SAR ATR development, the terminology and scope of this task have undergone significant evolutions. In early literature, \textit{detection} often referred only to the prescreening stage \cite{2008gaoguidetectionsurvey,el2016automaticSurvey,lan2020SARsurvey}. With the improvement of methods integration, \textit{detection} has gradually covered both the prescreening and discrimination stages \cite{li2023RScomprehensive}. To ensure the consistency and clarity of the discussion, this paper uniformly refers to both the process of extracting and screening candidate target regions as \textit{detection}, and the subsequent process of category inference as \textit{classification}. In conclusion, the SAR ATR task discussed in this paper refers to detecting the positions of potential targets in large-scene, single-channel, and static SAR images and then classifying their categories (Fig. \ref{definition and challenges of SAR ATRv2} (a)).

Based on the task definitions above, the objective of SAR ATR is to achieve reliable and efficient target detection and classification in complex operational environments. In this context, an effective SAR ATR framework should be evaluated from the following four aspects:

\textbf{(i) Accuracy:} Achieve high detection rates with low false-alarm rates, while maintaining reliable classification performance in cluttered scenes. In particular, the system should balance precision and recall and produce detection and classification results that are consistent with the underlying target categories.

\textbf{(ii) Adaptability and generalization:} Generalize to novel target categories, unseen imaging conditions, and heterogeneous sensors (e.g., different frequency bands or spatial resolutions). The model should also remain effective when labeled training data are limited, including few-shot and zero-shot settings.

\textbf{(iii) Robustness:} Maintain stable performance under target variations (e.g., pose changes, articulation, and geometric deformation), environmental effects (e.g., speckle noise, layover, and occlusion), and external interference such as adversarial perturbations or electronic jamming.

\textbf{(iv) Efficiency:} Meet the computational, memory, and power constraints of spaceborne and airborne platforms, while supporting low-latency inference for real-time or near-real-time applications.

\subsection{Core Challenges}

\begin{figure}[tb]
\centering
\includegraphics[width=0.45\textwidth]{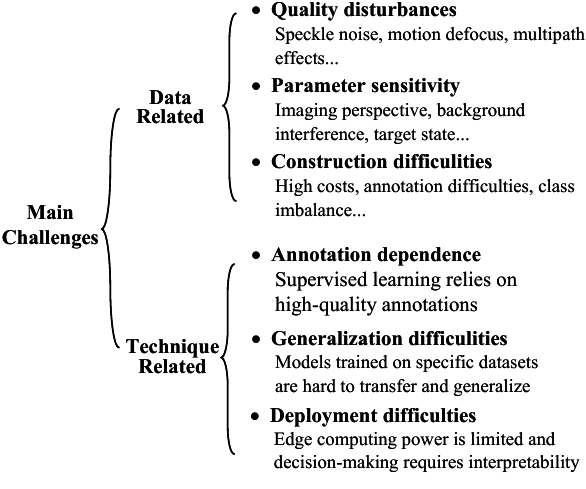}
\caption{Main Challenges of SAR ATR.}
\label{SAR_ATR_challenge}
 \vspace{-0.5cm}
\end{figure}

Despite continuous progress over the past decades, deploying SAR ATR systems in real-world operational environments remains challenging. Owing to the physical differences between microwave scattering and optical imaging, methods developed for computer vision cannot always be directly transferred to SAR without appropriate adaptation. As illustrated in Fig. \ref{SAR_ATR_challenge}, these challenges can be broadly categorized into data-related and technique-related aspects.

\textbf{\textit{1) Data-related Challenges:}} SAR data are subject to acquisition constraints, scene complexity, and limited annotation availability, all of which pose difficulties for data-driven ATR methods.

\textit{First, inherent imaging characteristics and background clutter complicate target representation.} Due to the coherent imaging mechanism of SAR, speckle is an inherent component of the observation process. Although speckle can reduce visual clarity and complicate feature extraction, it also reflects underlying scattering statistics and can be informative in applications such as PolSAR analysis and texture-based classification. In practical scenes, targets are often embedded in complex backgrounds (e.g., urban areas or dense vegetation), which may lead to a low signal-to-clutter ratio (SCR). In addition, SAR images can be affected by geometric effects such as layover, foreshortening, and shadow, as well as by multipath scattering and radio frequency interference (RFI) (Fig. \ref{definition and challenges of SAR ATRv2} (b)). Motion-induced defocusing and target wakes may further alter target appearance and spatial continuity.

\textit{Second, sensitivity to Extended Operating Conditions (EOCs).} The backscattering characteristics of a target are sensitive to multiple factors, including target attributes (e.g., articulation, configuration, and material), environmental conditions (e.g., occlusion and camouflage), and sensor parameters (e.g., grazing angle and resolution). Even relatively small changes in azimuth or depression angle may modify the dominant scattering centers of a target. As a result, intra-class variation can increase substantially, while inter-class separability may decrease, which makes robust recognition under EOCs difficult.

\textit{Third, the cost of annotation and data imbalance.} Acquiring measured SAR datasets is considerably more expensive than collecting optical imagery. As a result, available datasets are often limited in size, class-imbalanced, and long-tailed. In addition, annotation usually requires domain expertise and can be affected by weak scatterers, layover, occlusion, and visual similarity among categories, which may introduce omissions or labeling errors. Simulated data can partly alleviate data scarcity, but the domain gap between simulated and measured SAR data often limits their direct applicability.

\textit{Fourth, scale discrepancy and sparse target distribution.} Modern high-resolution SAR systems can cover wide geographic areas, whereas targets such as vehicles or ships usually occupy only a small fraction of the image (Fig. \ref{definition and challenges of SAR ATRv2} (b)). This spatial imbalance increases computational burden and makes small or subtle target signatures more difficult to distinguish from surrounding clutter.


\textbf{\textit{2) Technique-related Challenges:} }Although deep learning methods have achieved strong performance on standard benchmarks (e.g., MSTAR), their application to real-world SAR ATR remains limited by several methodological issues. Some of these issues are common to modern deep learning, whereas others are more closely related to the physical characteristics of SAR data and operational environments.

\textit{First, dependence on labeled data and the mismatch between optical pre-training and SAR representation.} Current ATR methods are still largely based on supervised learning and therefore depend on sufficient labeled training data. Because annotated SAR datasets are often limited, it is common to initialize models from optical pre-training (e.g., ImageNet \cite{deng2009imagenet}). While this strategy can improve optimization and data efficiency, it does not fully account for the complex-valued, coherent, and potentially multi-polarimetric nature of SAR observations. As a result, features transferred from optical imagery may not adequately capture SAR-specific scattering characteristics, which can limit adaptation to new targets, sensors, or operating conditions.

\textit{Second, limited generalization under EOCs and insufficient physical interpretability.} Models trained under Standard Operating Conditions (SOCs) often show degraded performance when evaluated under EOCs, such as changes in depression angle, target configuration, or sensor setting. One reason is that data-driven models may rely on correlations that are effective within a specific dataset but do not remain stable across operating conditions. In SAR, this issue is further complicated by the fact that image appearance is closely linked to scattering mechanisms and imaging geometry. Consequently, improving generalization requires not only more robust learning strategies but also representations that better reflect physically meaningful scattering structure.

\textit{Third, robustness to interference and deployment on resource-constrained platforms.} Practical SAR ATR systems may operate in dynamic electromagnetic environments and can be affected by both unintentional disturbances and intentional interference. Similar to other deep learning systems, SAR models can also be sensitive to adversarial perturbations; in addition, they may face SAR-relevant interference sources such as radio frequency interference or electronic countermeasures. At the same time, many operational applications require deployment on onboard or edge platforms, including satellites and unmanned aerial systems, where memory, computation, and power are limited. These constraints make it necessary to jointly consider robustness and computational efficiency in algorithm design.

\begin{table*}[!tb]
\setlength{\abovecaptionskip}{0.3cm} 
\centering
\footnotesize
\caption{\textbf{Summary of OPEN-SOURCE SAR target CLASSIFICATION datasets from the 1990s to the 2020s.} (Cls.: Number of target classes. Types: Number of target types. Img.: Number of images. Res.: Resolution. Pol.: Polarization. GF-3: Gaofen-3, S-1: Sentinel-1.)}
\label{table:atr_dataset}
\renewcommand\arraystretch{1.2}
\resizebox{0.99\linewidth}{!}{%
\begin{tabular}{|c|c|c|c|c|c|c|c|c|c|c|c|c|}
    \hline
        Dataset &  Year   & Link & Country &  Target &  Source  &  Band  &  Pol.  & Cls.  &  Types  &  Res.(m)  &  Img. Size  &   Img.   \\ \hline
        MSTAR \cite{MSTAR}& 1995  & \href{https://www.sdms.afrl.af.mil/index.php?collection=mstar}{link}& USA & vehicle & airborne & X & single & 8 & 10 & 0.3 & 128-193 & 14,557  \\ \hline
        CV Domes \cite{dungan2010civilian} & 2010  & \href{https://www.sdms.afrl.af.mil/index.php?collection=cv_dome}{link}& USA & vehicle& 3D simulation & X & quad & 3 & 10 & 0.3 & - & -  \\ \hline
        Gotcha \cite{dungan2012wide} & 2012 & \href{https://www.sdms.afrl.af.mil/index.php?collection=gotcha}{link}& USA & vehicle & 3D airborne & X & - & 7 & 13 & 0.3 & - & -  \\ \hline
        SARSIM \cite{malmgren2017improving} & 2017& \href{https://zenodo.org/records/573750}{link}& Denmark  & vehicle & simulation CAD & X & - & 7 & 14 & 0.1 & 139 & 21,168  \\ \hline
        OpenSARShip \cite{huang2018opensarship} & 2018& \href{https://opensar.sjtu.edu.cn/}{link}& China & ship & S-1 & C & dual & 16 & - & 2.7-22 & 9-445 & 26,679  \\ \hline
        SAMPLE \cite{lewis2019sar} & 2019 &\href{https://github.com/benjaminlewis-afrl/SAMPLE_dataset_public/tree/master}{link} &USA & vehicle & simulation & X & single & 7 & 10 & 0.3 & 128 & 2,690  \\ \hline
        FUSAR-Ship \cite{hou2020fusar} & 2020 &\href{https://radars.ac.cn/web/data/getData?dataType=FUSAR}{link} & China & ship & GF-3 & C & dual & 98 & - & 1.1-1.7 & 512 & 5,243  \\ \hline
        MATD \cite{wang2022multiangle} & 2022 & \href{https://www.radars.ac.cn/web/data/getData?newsColumnId=1c9a6287-4763-4f94-889e-156f50aca946}{link}& China & aircraft & airborne & Ku & - & 2 & 2 & - & 128 & 144  \\ \hline
        SAR-ACD \cite{sun2022scan} & 2022 & \href{https://github.com/AICyberTeam/SAR-ACD}{link}& China & aircraft& GF-3 & C & single & 2 & 6 & 1 & 32-200 & 3,032  \\ \hline
        ATRNet-STAR \cite{liu2025atrnet} & 2026 & \href{https://github.com/waterdisappear/ATRNet-STAR?tab=readme-ov-file}{link} &China & vehicle & airborne & X, Ku & quad & 21 & 40 & 0.12-0.15 & 128 & 194,324  \\ \hline
    \end{tabular}
}
 \vspace{-0.3cm}
\end{table*}

\begin{table*}[!tb]
\setlength{\abovecaptionskip}{0.1cm} 
\centering
\footnotesize
\caption{\textbf{Summary of OPEN-SOURCE SAR target DETECTION datasets from the 1990s to the 2020s.} (Cls.: Number of target classes. Img.: Number of images. Ins.: Number of instances. Res.: Resolution. Pol.: Polarization. GF-3: Gaofen-3, S-1: Sentinel-1. * and $\diamond$ represent horizontal and oriented target detection.)}
\label{table:detection_dataset}
\renewcommand\arraystretch{1.2}
\resizebox{0.99\linewidth}{!}{%
\begin{tabular}{|c|c|c|c|c|c|c|c|c|c|c|c|c|c|}
    \hline
        Dataset  & Year & Link & Country & Target &  Source  &  Band  &  Pol.  &   Cls.  &  Res.(m)  &  Img. Size  &  Img.  & Ins. & Ins./Img.  \\ \hline

*miniSAR \cite{miniSAR} & 2005 & \href{https://www.sandia.gov/radar/pathfinder-radar-isr-and-synthetic-aperture-radar-sar-systems/complex-data/}{link}& USA  & vehicle & airborne & Ku & - & 1 & 0.1 & 1638*2510 & 20 & - & -  \\ \hline

\multirow{2}{*}{*FARADSAR \cite{FARADSAR1,FARADSAR2}} & \multirow{2}{*}{2015}& \multirow{2}*{\href{https://www.sandia.gov/radar/pathfinder-radar-isr-and-synthetic-aperture-radar-sar-systems/complex-data/}{link}}& \multirow{2}{*}{USA}  & \multirow{2}{*}{vehicle} & \multirow{2}{*}{airborne} & \multirow{2}{*}{Ka,X} & \multirow{2}{*}{-} & \multirow{2}{*}{1} & \multirow{2}{*}{0.1} & 1300*580 & \multirow{2}{*}{412} &\multirow{2}{*}{ -} & \multirow{2}{*}{-}  \\ 
 & &&  &   &   &  &  &  &  &-1700*1850 &  &  &  \\ \hline
       \multirow{2}{*}{*SSDD \cite{li2017shipssdd}}  & \multirow{2}{*}{2017}&\multirow{2}*{\href{https://pan.baidu.com/s/1E8ixqK5AVfXc98UgQmpqaw}{link}} & \multirow{2}{*}{China} & \multirow{2}{*}{ship} & S-1,RadarSat-2 & \multirow{2}{*}{C/X} & HH,VV, & \multirow{2}{*}{1} & \multirow{2}{*}{1-15} & \multirow{2}{*}{160-668} & \multirow{2}{*}{1,160} & \multirow{2}{*}{2,456} & \multirow{2}{*}{2.12}  \\ 

 & &&  &   & TerraSAR-X &  &VH,HV &  &  & &  &  &  \\ \hline

        *AIR-SARSHIP1.0 \cite{xian2019airsarship1} & 2019 & \href{https://aistudio.baidu.com/datasetdetail/112784}{link}& China  & ship & GF-3 & C & Single & 1 & 1, 3 & 3000 & 31 & 461 & 14.87  \\ \hline
        *AIR-SARSHIP2.0 \cite{xian2019airsarship1}  & 2019  & \href{https://aistudio.baidu.com/datasetdetail/112784}{link}& China & ship & GF-3 & C & Single & 1 & 1, 3 & 1000 & 300 & 2,040 & 6.8  \\ \hline
        *SAR-SHIP-DATASET \cite{wang2019sarshipdataset} & 2019& \href{https://github.com/CAESAR-Radi/SAR-Ship-Dataset}{link}& China  & ship & S-1,GF-3 & C & Single, Dual, Full & 1 & 3-25 & 256 & 39,729 & 47,416 & 1.2  \\ \hline
       *LS-SSDD-v1.0 \cite{zhang2020lsssdd}& 2020& \href{https://github.com/zhangfx123/LS-SSDD-v1.0-OPEN}{link}& China & ship & S-1 & C & VV,VH & 1 & 5-20 & 800 & 9000 & 6015 & 0.67  \\ \hline
       
       \multirow{2}{*}{*HRSID \cite{wei2020hrsid}}  &  \multirow{2}{*}{2020} & \multirow{2}{*}{\href{https://github.com/chaozhong2010/HRSID}{link}}& \multirow{2}{*}{China} &  \multirow{2}{*}{ship} &  S-1B,TerraSAR-X, &  \multirow{2}{*}{C/X} &  \multirow{2}{*}{HH,HV,VV} &  \multirow{2}{*}{1} &  \multirow{2}{*}{0.5-3} &  \multirow{2}{*}{800}
       &  \multirow{2}{*}{5,604} &  \multirow{2}{*}{16,951} &  \multirow{2}{*}{3.02}  \\

 &  &  & &  &TanDEMX & & & &  & &  &  &   \\ \hline

    \multirow{2}{*}{$\diamond$ SRSDD-v1.0 \cite{lei2021srsdd}}& \multirow{2}{*}{2021} & \multirow{2}{*}{\href{https://github.com/HeuristicLU/SRSDD-V1.0}{link}}& \multirow{2}{*}{China}  & \multirow{2}{*}{ship} & \multirow{2}{*}{GF-3} & \multirow{2}{*}{C} & \multirow{2}{*}{HH,VV} & \multirow{2}{*}{1} & \multirow{2}{*}{1} & \multirow{2}{*}{1024} & \multirow{2}{*}{666} & \multirow{2}{*}{2,884} & \multirow{2}{*}{4.33}  \\ 
    
      &   &  & &   &   &   &  &  (6 sub) &   &   &   &   &    \\ \hline
    
        \multirow{2}{*}{$\diamond$ *Official SSDD \cite{zhang2021sarssdd}} & \multirow{2}{*}{2021} &  \multirow{2}{*}{\href{https://github.com/TianwenZhang0825/Official-SSDD}{link}} &\multirow{2}{*}{China} & \multirow{2}{*}{ship} & S-1,RadarSat-2 & \multirow{2}{*}{C/X} & HH,VV, & \multirow{2}{*}{1} & \multirow{2}{*}{1-15} & \multirow{2}{*}{160-668} & \multirow{2}{*}{1,160} & \multirow{2}{*}{2,456} & \multirow{2}{*}{2.12}  \\ 

 &  & & & & TerraSAR-X &  &VH,HV &  &  & &  &  &  \\ \hline

        $\diamond$ *DSSDD \cite{hu2021dssdd} & 2021 & \href{https://github.com/liyiniiecas/A_Dual-polarimetric_SAR_Ship_Detection_Dataset}{link}& China & ship & S-1 & C & VV,VH & 1 & 9,14 & 256 & 1,236 & 3,540 & 2.86  \\ \hline

           $\diamond$ RSDD-SAR \cite{xu2022rsdd} & 2022& \href{https://github.com/makabakasu/RSDD-SAR-OPEN }{link}&China & ship & GF-3, TerraSAR-X & C/X & HH,HV & 1 & 2-20 & 512 & 7,000 & 10,263 & 14.66  \\ \hline

    *SADD \cite{zhang2022saddsefepnet}& 2022& \href{https://github.com/hust-rslab/SAR-aircraft-data}{link}& China & aircraft & TerraSAR-X & X & HH & 1 & 0.5-3 & 224 & 2,966 & 7,835 & 2.64  \\ \hline
           
    \multirow{2}{*}{*MSAR \cite{xia2022crtranssarMSAR}} & \multirow{2}{*}{2022}&\multirow{2}{*}{\href{https://radars.ac.cn/web/data/getData?dataType=MSAR}{link}}& \multirow{2}{*}{China} & aircraft, ship, & \multirow{2}{*}{HISEA-1} & \multirow{2}{*}{C}& HH,HV & \multirow{2}{*}{4} & \multirow{2}{*}{ 1} & \multirow{2}{*}{ 256-2048} & \multirow{2}{*}{ 28,449} & \multirow{2}{*}{ 60,396} & \multirow{2}{*}{ 2.12}  \\ 
-

    &  & &  & bridge, oil tank & &  & VH,VV &  &  &  &  &  &   \\ \hline
    
      \multirow{2}{*}{*SAR-AIRcraft1.0 \cite{zhirui2023sarcraft}} & \multirow{2}{*}{2023} & \multirow{2}{*}{\href{https://aistudio.baidu.com/datasetdetail/312407}{link}}& \multirow{2}{*}{China} & \multirow{2}{*}{aircraft} & \multirow{2}{*}{GF-3} & \multirow{2}{*}{C} & \multirow{2}{*}{Uni-polar} & \multirow{2}{*}{1} & \multirow{2}{*}{1} & \multirow{2}{*}{800-1500} & \multirow{2}{*}{4,368} & \multirow{2}{*}{16,463} & \multirow{2}{*}{3.77}  \\ 

             &   &  & &   &   &   &  &  (7 sub) &   &   &   &   &    \\ \hline
      
       *SIVED \cite{lin2023sived} & 2023 & \href{https://github.com/CAESAR-Radi/SIVED}{link}&China & vehicle &airborne & Ka,Ku,X & VV/HH & 1 & 0.1, 0.3 & 512 & 1,044 & 12,013 & 11.51  \\ \hline

        \multirow{2}{*}{$\diamond$ *OGSOD\cite{wang2023ogsod}} & \multirow{2}{*}{2023} & \multirow{2}{*}{\href{https://github.com/mmic-lcl/Datasets-and-benchmark-code}{link}}& \multirow{2}{*}{China}  & bridge, oil tank, & \multirow{2}{*}{GF-3} & \multirow{2}{*}{C} & \multirow{2}{*}{VV/VH} & \multirow{2}{*}{3} & \multirow{2}{*}{3} & \multirow{2}{*}{256} & \multirow{2}{*}{18,331} & \multirow{2}{*}{48,589} & \multirow{2}{*}{2.65} \\ 
  &  & &  &  harbour & &  &  &  &  &  &  &  &   \\ \hline

           \multirow{3}{*}{*SARDet-100k \cite{li2024sardet100}}  & \multirow{3}{*}{2024} & \multirow{3}{*}{\href{https://github.com/zcablii/SARDet_100K}{link}}& \multirow{3}{*}{China} &  aircraft, ship,&TerraSAR-X,TanDEMX &\multirow{2}{*}{Ka,Ku,}& \multirow{2}{*}{HH,HV,} & \multirow{3}{*}{6}& \multirow{3}{*}{0.1-25} & \multirow{3}{*}{512} &\multirow{3}{*}{116,598} & \multirow{3}{*}{245,653} & \multirow{3}{*}{2.11}  \\ 
           
               &   &  &   & bridge, oil tank,  &RadarSat-2,Airborne  &   &   &   &   &   &   &   &    \\
            &   &    &  & vehicle, harbour  &HISEA-1,GF-3,S-1B &  X,C & VH,HV  &   &   &   &   &   &    \\
            \hline

               \multirow{3}{*}{$\diamond$ FAIR-CSAR \cite{wu2024faircsar}}  & \multirow{3}{*}{2024} & \multirow{3}{*}{\href{https://radars.ac.cn/web/data/getData?dataType=FAIR_CSAR}{link}}& \multirow{3}{*}{China} &  aircraft, ship,&\multirow{3}{*}{GF-3} &\multirow{3}{*}{C}& \multirow{2}{*}{HH,HV,} & \multirow{3}{*}{5}& \multirow{3}{*}{1-5} & \multirow{3}{*}{1024} &\multirow{3}{*}{106,672} & \multirow{3}{*}{349,002} & \multirow{3}{*}{3.27}  \\ 
           
               &   &  &   & bridge, oil tank,  &  &   &   &   &   &   &   &   &    \\ 
               &   &    &  & tower crane  &  &    & VH,HV  &  (22 sub) &   &   &   &   &    \\
               \hline
     \multirow{3}{*}{$\diamond$ RSAR \cite{zhang2025rsar}}  & \multirow{3}{*}{2025} & \multirow{3}{*}{\href{https://github.com/zhasion/RSAR}{link}}& \multirow{3}{*}{China} &  aircraft, ship,&TerraSAR-X,TanDEMX &\multirow{2}{*}{Ka,Ku,}& \multirow{2}{*}{HH,HV,} & \multirow{3}{*}{6}& \multirow{3}{*}{0.1-25} & \multirow{3}{*}{512} &\multirow{3}{*}{95,842} & \multirow{3}{*}{183,534} & \multirow{3}{*}{1.91}  \\ 
           
               &   &  &   & bridge, tank,  &RadarSat-2,Airborne  &   &   &   &   &   &   &   &    \\
            &   &    &  & car, harbour  &HISEA-1,GF-3,S-1B &  X,C & VH,HV  &   &   &   &   &   &    \\
            \hline

    \end{tabular}
    }
     \vspace{-0.2cm}
\end{table*}

\begin{figure}[ht]
\centering
\includegraphics[width=0.45\textwidth]{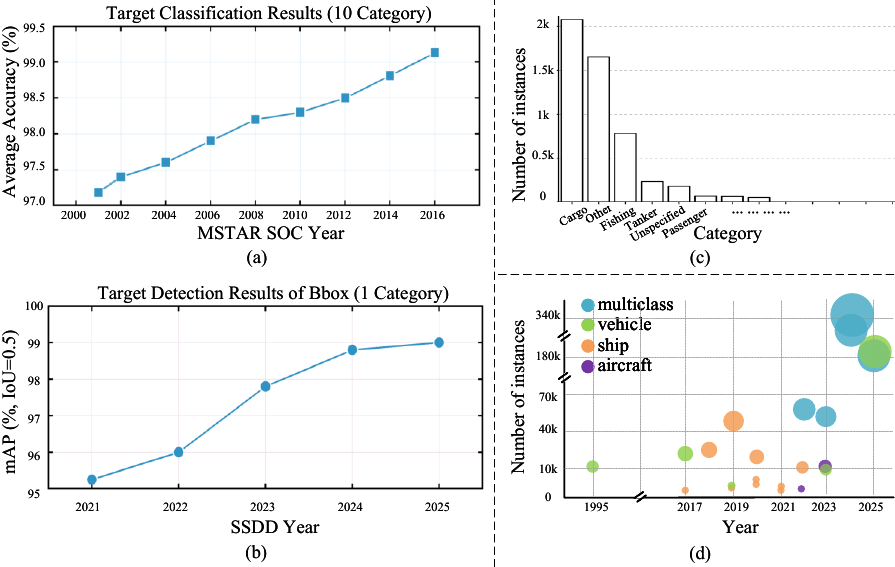}
\caption{Development status and challenges of SAR ATR datasets. (a) Annual average classification accuracy on MSTAR  \cite{keydel1996mstar} under SOC. (b) Annual variation of mAP on Bbox SSDD  \cite{li2017shipssdd,zhang2021sarssdd}. (c)  Instances count distribution across different categories in FuSAR-Ship  \cite{hou2020fusar}, presenting a significant long-tailed phenomenon. (d) Scale (instances) and category coverage of released SAR ATR datasets in recent years, reflecting the trend of datasets developing toward larger scales and more categories.}
\label{challenges of datasets}
\end{figure}

\subsection{Datasets and Evaluation}
\label{sec:DE}

\textbf{\textit{1) Datasets:}} Constructing larger datasets with smaller biases is crucial for developing advanced detection and recognition algorithms. Over the past five decades, the development history of SAR ATR datasets has itself been a technical history that drives the evolution of paradigms in this field. TABLE \ref{table:atr_dataset} and TABLE \ref{table:detection_dataset} present the currently available open-source classification and detection datasets, along with official download links.

\textbf{\textit{2) Evaluation Metrics:}} How do we evaluate the accuracy of SAR ATR systems? The answer to this question may vary over time. In the early research on detection, there were no widely accepted metrics for evaluating detection accuracy. For example, in early studies on ATR systems \cite{novak1993performance}, Novak used the probability of detection for uncamouflaged and camouflaged targets and confusion matrices to assess the classification accuracy of the system. Later, ATR methods typically categorized detection results into correct detections (where targets are correctly identified) and false alarms (where non-target objects are mistakenly classified as targets). The key performance indicators for these methods include the probability of detection (PD) and the probability of false alarm (PFA). Particularly in Constant False Alarm Rate (CFAR) detectors \cite{hansen1980detectabilityGOCFAR, gandhi2002analysisTMCFAR}, maintaining a constant false alarm rate under various conditions is crucial. In recent years, the most commonly used detection evaluation metrics are accuracy and AP. AP is defined as the average detection precision across different recall rates, typically class-specifically \cite{liu2020deep}. The mean AP (mAP) across all classes is typically used as the final performance indicator. More details are summarized in TABLE \ref{Tab:Metrics}, and further details are provided in \cite{liu2020deep} and \cite{slesinski2025review}.

\textbf{\textit{3) Dataset Limitations and Development Trends:}} Classical datasets such as MSTAR \cite{keydel1996mstar} and SSDD \cite{li2017shipssdd, zhang2021sarssdd} have played an important role in the development of SAR target classification and detection. As shown in Fig. \ref{challenges of datasets} (a) and (b), recent methods have achieved very high performance on these benchmarks, suggesting that evaluation on them is approaching saturation. This trend indicates that, although these datasets remain valuable for standardized comparison, their limited scale, category diversity, imaging conditions, and scene complexity may restrict their ability to reflect real-world SAR ATR challenges. In addition, some public datasets exhibit long-tailed category distributions and inter-class imbalance (Fig. \ref{challenges of datasets} (c)), which may further affect the assessment of model generalization in practical scenarios. In response, recent studies have increasingly focused on constructing larger and more diverse datasets with richer categories and more detailed annotations (Fig. \ref{challenges of datasets} (d)). Overall, this development reflects a shift from relatively small and controlled benchmarks toward large-scale, multi-scene, and more challenging datasets that are better suited for evaluating SAR ATR methods in practical applications.

\begin{table}[!t]
\setlength{\abovecaptionskip}{0.1cm} 
\setlength{\belowcaptionskip}{-0.1cm}
\caption{Summary of commonly used metrics for evaluating SAR ATR methods.}\label{Tab:Metrics}
\centering
\renewcommand{\arraystretch}{1.2}  
\setlength\arrayrulewidth{0.2mm}    
\setlength\tabcolsep{3pt}           
\resizebox*{8.8cm}{!}{  
\begin{tabular}{|c|c|l|l|}  
\hline
\footnotesize Metric  & \footnotesize Meaning  &  \multicolumn{2}{c|}{\footnotesize Definition and Description} \\
\hline
\footnotesize TP  & \footnotesize \shortstack[c]{True \\ Positive}  &  \multicolumn{2}{l|}{\footnotesize A true positive detection.}\\
\hline
\footnotesize FP  & \footnotesize \shortstack[c]{False \\ Positive}  &  \multicolumn{2}{l|}{\footnotesize A false positive detection.} \\
\hline
\footnotesize FN  & \footnotesize \shortstack[c]{False \\ Negative}  &  \multicolumn{2}{l|}{\footnotesize A false negative detection.} \\
\hline

\footnotesize TN  & \footnotesize \shortstack[c]{True \\ Negative}  &  \multicolumn{2}{l|}{\footnotesize A true negative detection.} \\
\hline

\footnotesize Acc  & \footnotesize \shortstack[c]{Accuracy \\ Rate}  &  \multicolumn{2}{l|}{\footnotesize $Accuracy = \frac{TP+TN}{TP+TN+FP+FN}$.} \\
\hline
\footnotesize FAR  & \footnotesize \shortstack[c]{False \\ Alarm \\ Rate}  &  \multicolumn{2}{l|}{\footnotesize  $FAR = \frac{FP}{TN+FP}$. } \\
\hline

\footnotesize F1 & \footnotesize \shortstack[c]{F1-score}  &  \multicolumn{2}{l|}{\footnotesize \shortstack[l] F1-score = $\frac{2 \cdot \text{Precision} \cdot \text{Recall}}{\text{Precision} + \text{Recall}}$.} \\
\hline

& \multirow{3}{*}{\footnotesize \shortstack[c]{mean \\ Average \\ Precision}}  &  \multicolumn{2}{l|}{\footnotesize $\bullet AP$: mAP averaged over ten IOUs: $\{0.5:0.05:0.95\}$;} \\
\cline{3-4}  
&  &  \multicolumn{2}{l|}{\footnotesize $\bullet AP^{\textrm{IOU}=0.5}$: mAP at IOU=0.50;} \\
\cline{3-4}
\footnotesize mAP &  &  \multicolumn{2}{l|}{\footnotesize $\bullet AP^{\textrm{IOU}=0.75}$: mAP at IOU=0.75 (strict metric);} \\
\hline
\footnotesize mAR  & \footnotesize \shortstack[c]{Average \\ Recall}  &  \multicolumn{2}{l|}{\footnotesize \shortstack[l]{The maximum recall given a fixed number of detections per image, \\ averaged over all categories and IOU thresholds.}} \\
\hline

\end{tabular}
}
 \vspace{-0.3cm}
\end{table}

%% file: Revision_noblue_section/HistoryOfSARATR.tex
\section{History of SAR ATR}
\label{sec:history}

\begin{figure*}[htbp!]
\centering
\includegraphics[width=0.9\textwidth]{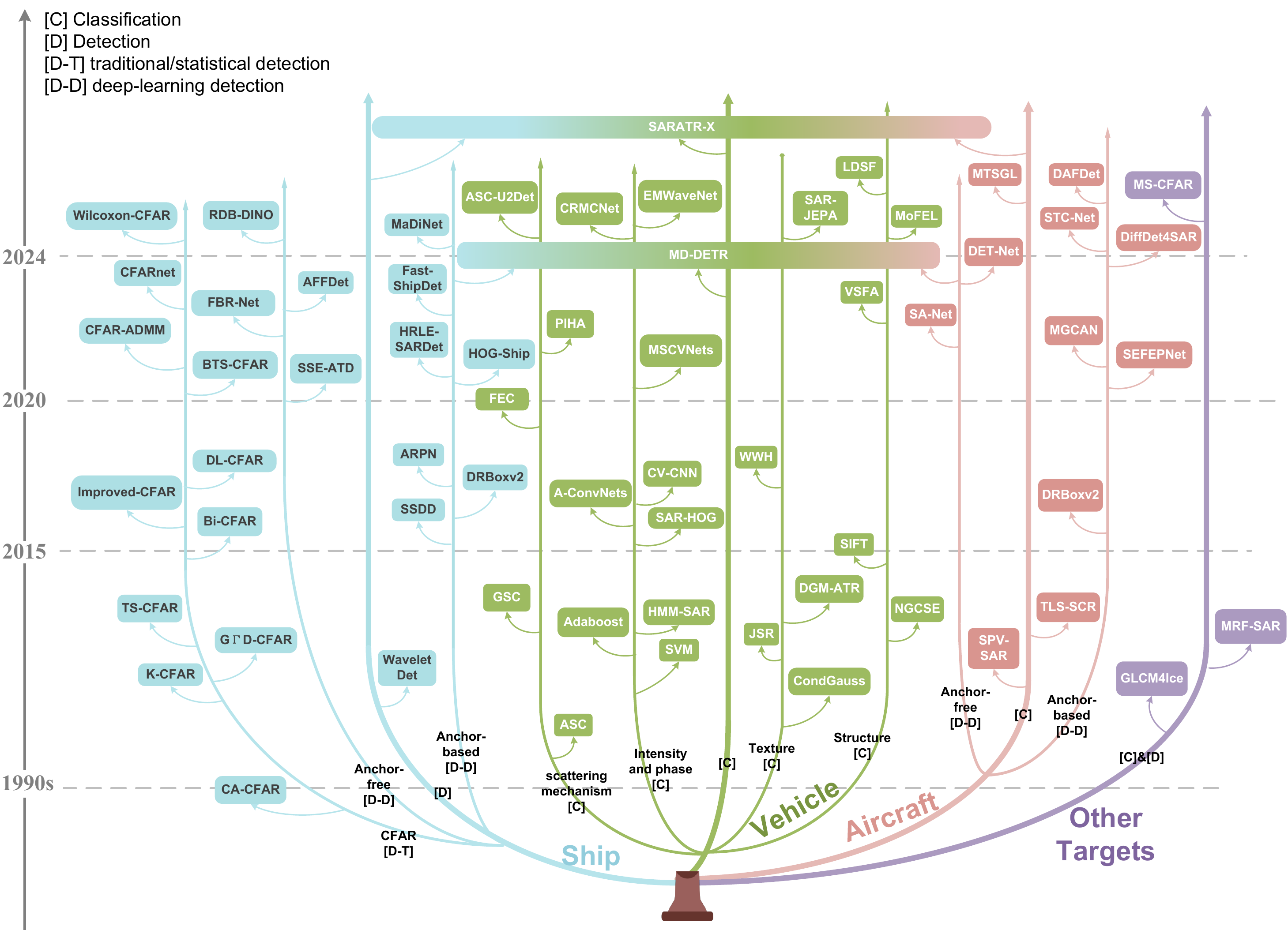}
\caption{An evolutionary tree of SAR ATR technology from the 1990s to the present, organized into four primary branches based on target types: ships, vehicles, aircraft, and other targets. Branch nodes represent landmark methodologies, connecting lines indicate technological inheritance and innovation, while cross-branch linkages signal the emergence of generalizable models. The figure highlights representative milestone methods rather than providing an exhaustive method list. The corresponding original references of the method nodes are provided in the detailed discussions of Sections IV and V. Several observations can be drawn from this evolutionary tree. (1) Early approaches (pre-2015) predominantly employed handcrafted features and statistical modeling, exemplified by CFAR variants focused on ship detection. The post-2015 period saw deep learning becoming mainstream through architectures like A-ConvNets \cite{chen2016target} and CV-CNN \cite{zhang2017complexCVCNN}, though most remained target-specific. Since 2020, generalist models such as MD-DETR \cite{liu2024MDDETR} and SARATR-X \cite{li2025saratrx} have demonstrated cross-target generalization capabilities. (2) Branch-specific development patterns. The ship detection branch exhibits the densest node distribution, reflecting high research maturity and methodological diversity. Vehicle detection, although emerging later, demonstrates an accelerating growth trajectory. Aircraft recognition remains heavily reliant on structural and scattering feature modeling. (3) Three key evolutionary trends in SAR ATR. A distinct shift from handcrafted feature engineering toward data-driven learning paradigms. A transition from single-target detection toward multi-target generalization capabilities. Increasing integration of physics-inspired methodologies with data fusion and model-driven frameworks. Considering the rapid development of SAR ATR, we share the source file of research in this free and encourage readers to make incremental updates at \href{https://github.com/JoyeZLearning/SAR-ATR-From-Beginning-to-Present}{https://github.com/JoyeZLearning/SAR-ATR-From-Beginning-to-Present}.} 
\label{SAR_ATR_tree}
 \vspace{-0.3cm}
\end{figure*}

Over the past 50 years, the development of SAR ATR has centered on the problem of target representation, while its application scope has gradually expanded from controlled target recognition to more diverse classification and detection scenarios. This evolution is reflected in the methodological tree (Fig. \ref{SAR_ATR_tree}), which uses target types as branches and method nodes to illustrate technological inheritance, innovation, and generalization. Fig. \ref{History of SAR ATR} further summarizes representative milestones in this process. Rather than presenting an exhaustive list of methods at a single technical level, the figure is organized around the two core SAR ATR tasks, namely classification and detection. Within each task branch, we highlight historically influential milestone events, including benchmark datasets, theoretical or statistical modeling advances, feature representation methods, classifiers or detectors, and later deep network architectures. Based on the dominant target representation paradigm, the development of SAR ATR can be broadly divided into four stages: understanding physics, designing features, learning features, and understanding-and-learning features. The transitions between these stages were mainly driven by the maturation of SAR imaging and scattering theory, the availability of benchmark datasets, advances in computing and machine learning techniques, and, more recently, the demand for improved generalization, robustness, and physical interpretability in practical applications.

\subsection{Understanding Physics: Theoretical Foundations and Statistical Modeling (1970s–1990s)}
Early research focused primarily on physical mechanism modeling and statistical theory, with the goal of establishing the foundations of SAR imaging and target scattering analysis. In 1951, Carl Wiley proposed the principle of Doppler Beam Sharpening (DBS) and described the frequency-domain conditions for synthetic aperture image formation \cite{wiley1985synthetic}. Rihaczek later established an early theoretical connection between electromagnetic target properties and recognition feasibility \cite{1969Principles}. Harger further standardized SAR imaging geometry, providing a reproducible basis for subsequent analysis and interpretation \cite{harger1971synthetic}. 

With the launch of Seasat-A in 1978, increasing access to measured SAR data promoted the statistical modeling of coherent imaging phenomena and background clutter. Representative studies included the speckle model \cite{arsenault1976speckleModel}, K-distribution \cite{jakeman1976KDistribution}, product model \cite{ward1981productmodel}, and texture analysis methods \cite{ulaby1986textural,burl1989texture}. In parallel, target detection research developed along a statistical decision-theoretic path. Cell-Averaging Constant False Alarm Rate (CA-CFAR) \cite{finn1966adaptiveCACFAR,hm1968adaptiveCACFAR} transformed the Neyman--Pearson criterion into an adaptive thresholding algorithm, enabling automatic detection under a controlled false-alarm rate. Novak \textit{et al.} \cite{novak1989studies} further combined clutter covariance estimation with multi-polarization channel fusion, providing an important basis for target detection in cluttered environments.

Overall, this stage established the physical and statistical foundations of SAR ATR. However, these approaches relied mainly on explicit modeling and hypothesis testing, which offered limited flexibility for representing fine-grained target variability. This limitation motivated the subsequent shift toward handcrafted feature design.

\subsection{Designing Features: Handcraft Feature Engineering (1990s–2010s)}
This stage marked the transition from establishing theoretical foundations to explicitly designing discriminative target representations for recognition. It was driven by the maturation of SAR imaging theory, the increasing availability of measured data, and the emergence of benchmark datasets and standardized evaluation protocols. During this period, SAR ATR research expanded from target detection toward finer-grained recognition and classification, with handcrafted feature engineering becoming the dominant paradigm. 

A major line of work in this stage focused on characterizing SAR targets from multiple complementary perspectives. Physical features, represented by Attributed Scattering Center (ASC) parameters \cite{potter1997attributed}, directly described the electromagnetic scattering mechanisms of targets. Statistical features modeled regional scattering behavior using parametric distributions such as the G$_{0}$ distribution \cite{frery1997model} and Fisher distribution \cite{tison2004newMK}. Structural features, including wavelet transform \cite{tello2005WaveletDet} and Gray-Level Co-occurrence Matrix (GLCM) \cite{holmes1998textural}, were widely used to capture target geometry and texture patterns. The release of the MSTAR dataset in 1996 and the subsequent SOC/EOC evaluation protocols \cite{ross1998standard} provided a standardized basis for method development and comparison. On this foundation, traditional machine learning methods such as SVM \cite{zhao2002SVMATR}, AdaBoost \cite{sun2007adaboost}, sparse representation (JSR) \cite{zhang2012multiJSR}, and SIFT-SAR \cite{dellinger2014sarSIFT} were introduced to improve the discriminative ability and robustness of handcrafted descriptors. In parallel, SAR target detection also evolved beyond basic thresholding. While CFAR-based methods continued to be refined \cite{leng2015bilateralCFAR,dai2016modifiedCFAR}, methods incorporating structural and contextual cues, such as Radon transform \cite{rey1990application}, morphological filtering \cite{eldhuset2002automatic}, edge detection \cite{oliver1996optimum}, and Markov Random Field (MRF) modeling \cite{tupin2002MRFDet}, gradually moved detection from pixel-level decisions toward more structured scene interpretation.

The defining characteristic of this stage was the combination of domain-expert-designed features and shallow learning algorithms. This paradigm substantially improved SAR ATR performance under controlled conditions, but it also depended heavily on prior knowledge and manual feature design, which limited scalability and adaptability. These limitations later motivated the transition toward data-driven representation learning.

\subsection{Learning Features: Data-Driven End-to-End Learning (2010s–2020s)}

This stage was characterized by the broad introduction of deep learning techniques \cite{krizhevsky2012imagenet}, whose central idea is to learn hierarchical feature representations directly from raw SAR images, thereby reducing reliance on manually designed descriptors. The transition to this stage was enabled by increased computing capability, progress in optimization algorithms, and the growing influence of large-scale representation learning in computer vision. 

Early studies \cite{ding2016CNNATR, chen2016target} demonstrated the effectiveness of convolutional neural networks (CNNs) for SAR ATR on benchmark datasets such as MSTAR. Compared with handcrafted pipelines, CNN-based models provided a more unified framework for jointly learning feature extraction and classification. Subsequent work explored how SAR-specific characteristics could be incorporated into deep architectures. For example, the Complex-Valued CNN (CV-CNN) \cite{zhang2017complexCVCNN} introduced phase information from complex SAR data into the end-to-end learning process, improving the completeness of feature representation. Later, the FEC framework \cite{zhang2020fec} fused electromagnetic scattering center features with CNN deep features through discriminant correlation analysis, illustrating the complementarity between physical priors and learned representations. For target detection, the release of datasets such as SSDD \cite{li2017shipssdd} enabled deep detection frameworks to be adapted to SAR imagery. General detectors, such as Faster R-CNN \cite{Fasterrcnn2017}, were progressively modified to better match SAR characteristics, leading to specialized designs involving attention mechanisms \cite{cui2019dapn}, rotated anchors \cite{an2019drboxV2}, and multi-scale feature fusion \cite{ai2019tcsJcfar}. These developments significantly improved detection accuracy and broadened the applicability of deep models to more complex scenes.

Overall, this stage demonstrated the effectiveness of end-to-end feature learning and established deep learning as the dominant paradigm in SAR ATR. However, the dependence of deep models on large-scale labeled data and their limited physical interpretability became increasingly evident in practice. These limitations motivated the recent shift toward more integrated physics-data dual-driven approaches.

\subsection{Understanding and Learning: Physics-Data Dual-Driven Fusion (2020s–Present)}

Recent SAR ATR research has increasingly moved toward the integration of physics-guided and data-driven learning. This transition has been motivated by the limitations of purely data-driven models in terms of generalization, robustness, and interpretability, especially under EOCs, cross-domain transfer, and resource-constrained deployment.

On the data side, the construction of larger-scale and more diverse datasets \cite{wu2024faircsar,zhang2025rsar} has provided an important foundation for training and evaluating more generalized models. On the model side, architectures such as Vision Transformers and state space models \cite{zhou2025madinet} have expanded the design space of SAR representation learning. At the same time, physics-aware modules, including physics-prior attention mechanisms \cite{huang2022piha, li2021CAMCNN} and diffusion-based frameworks \cite{zhou2024diffdet4sar, zhou2025madinet}, have been introduced to incorporate electromagnetic scattering principles into deep networks more explicitly. The learning paradigm has also evolved beyond purely supervised training. Self-supervised learning and cross-domain pre-training \cite{li2024sarjepa,li2025saratrx} increasingly exploit large amounts of unlabeled data to reduce annotation dependence and improve transferability under few-shot or domain-shift settings. As a result, tasks such as detection, recognition, and segmentation can now be adapted more flexibly within unified foundation-model-style frameworks \cite{zhang2024earthgpt, li2025tacmt}, showing encouraging scalability across tasks and scenarios.

In this sense, SAR ATR has evolved from physics-driven modeling, to handcrafted feature engineering, to data-driven representation learning, and now toward physics-data dual-driven fusion. This emerging paradigm provides a promising direction for building SAR ATR systems that are not only accurate, but also more interpretable, robust, and practically deployable.

%% file: Revision_noblue_section/EvolutionOfDetection.tex
\section{Evolution of SAR Target Detection}
\label{sec:detection}

\begin{figure}[ht]
\centering
\includegraphics[width=0.495\textwidth]{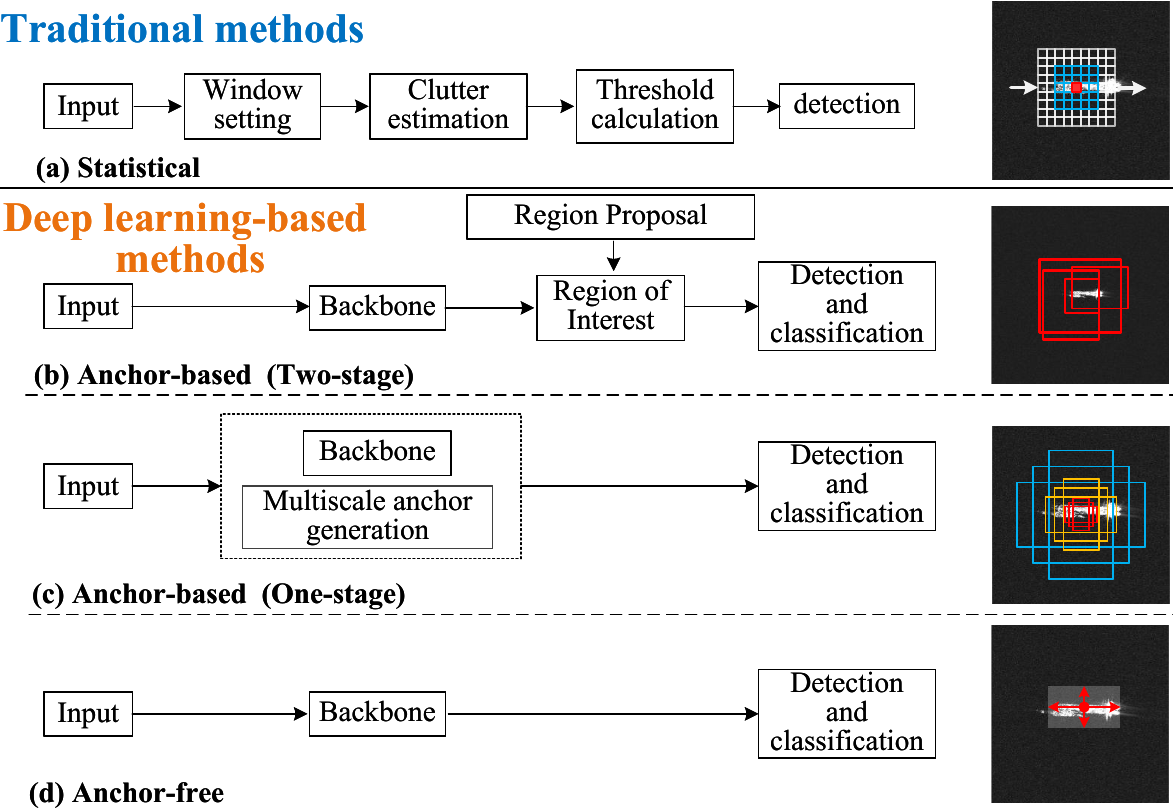}
\caption{Overview of key steps between traditional and deep learning-based methods for SAR target detection. (a) Statistical methods. (b) Anchor-based (two-stage) methods. (c) Anchor-based (one-stage) methods. (d) Anchor-free methods.}
\label{detection comparision high level}
     \vspace{-0.5cm}
\end{figure}

The evolution of SAR target detection has transitioned from model-driven exploitation of physical priors to data-driven representation learning. \textbf{Its core challenge remains: robustly and accurately isolating targets from strong speckle and complex backgrounds while suppressing false alarms.}

\subsection{Traditional Methods for SAR Target Detection}

\textbf{\textit{1) Statistical Feature-based Methods:}} Traditional SAR target detection formulates the task as a statistical hypothesis test, and the constant false alarm rate (CFAR) detector family is the most widely adopted implementation of this principle \cite{alexandre2024shipreview}, such as Improved-CFAR \cite{li2017improvedSPCFAR} and BiTS-CFAR \cite{ai2021robustBTSCFAR}. CFAR partitions the local background with a sliding window and adaptively sets the detection threshold from the clutter distribution. This strategy maintains a constant false alarm rate in complex and time-varying electromagnetic environments (Fig. \ref{detection comparision high level} (a)). This physics-driven approach models inherent SAR phenomena, such as multiplicative speckle and non-stationary clutter, using quantifiable statistical distributions (e.g., Rayleigh, Weibull, K, and generalized gamma). Consequently, CFAR detectors are categorized into parametric CFAR and background modeling schemes.

\textit{(i) Window Setting-based Parametric CFAR:} Originating from cell averaging CFAR (CA-CFAR) \cite{finn1966adaptiveCACFAR,hm1968adaptiveCACFAR,weiss2007ModifiedCACFAR}, the parameterized branch has produced OS-CFAR \cite{rohling2007radarOSCFAR}, SO-CFAR \cite{trunk1978rangeSOCFAR}, GO-CFAR \cite{hansen1980detectabilityGOCFAR}, TM-CFAR \cite{gandhi2002analysisTMCFAR} and others \cite{smith2002intelligentVICFAR}. Each variant applies a distinct nonlinear transformation to the background window to survive nonhomogeneous scenes. SO-CFAR \cite{trunk1978rangeSOCFAR} minimizes the power estimate between leading and lagging windows to resolve closely spaced targets. GO-CFAR \cite{hansen1980detectabilityGOCFAR} takes the maximum estimate to reduce masking from interfering targets. OS-CFAR \cite{rohling2007radarOSCFAR} replaces the sample mean with a ranked statistic, which yields robustness near the clutter edges. TM-CFAR \cite{gandhi2002analysisTMCFAR} symmetrically censors extreme samples, trading a slight loss in signal-to-noise ratio for significantly improved adaptability to environmental transitions. However, these detectors face an intrinsic trade-off in selecting the size of the background window. A large window tends to straddle heterogeneous regions and introduces contaminated samples, biasing the clutter model. A small window provides insufficient samples, inflating the variance of the estimated parameters and causing erratic thresholds. Although Ratio-CFAR \cite{touzi1988statisticalRatioCFAR} reduces false alarms induced by speckle and BLUE-CFAR \cite{di2005BLUEcfar} employs a Weibull-Gumbel transform to account for self-shadowing of extended targets, such refinements do not overcome the fundamental limitations imposed by model mismatch and poor scene adaptability.

\textit{(ii) Clutter Estimation-based CFAR:} To cope with non Gaussian and spatially inhomogeneous clutter, refined CFAR schemes have been introduced that assume specific complex distributions \cite{goldstein2007falseLOGT}. K-CFAR \cite{kuttikkad1994KCFAR} models spiky sea clutter by a K distribution and employs a guard band reference window to reduce target leakage into the background estimate. G$\Gamma$D-CFAR \cite{qin2012GFDcfar} derives a closed-form threshold for high-resolution sea clutter by replacing the conventional distribution with the generalized Gamma distribution. Similar strategies adopt the generalised gamma (TS-CFAR) \cite{tao2015TSCFAR} or other flexible distributions, including Bi-CFAR \cite{leng2015bilateralCFAR} and Wilcoxon-CFAR \cite{meng2025wilcoxonCFAR} to capture the complex scenarios encountered in ground and sea regions. Yet the core difficulty remains that a single parametric form cannot accommodate the abrupt statistical transitions present at urban edges, within densely packed harbors, or across mountainous terrain, and the resulting model mismatch continues to degrade detection performance in complex scenes.

\textit{(iii) Others:} Beyond pixel-level grayscale statistics like CFAR, some works transform SAR images into multiscale or wavelet domains, leveraging differences in high-frequency energy, coefficient distribution, or correlation between targets and backgrounds to achieve detection.  Tello \textit{et al.} \cite{tello2005WaveletDet} leveraged discrete wavelet transforms to enhance multiscale discontinuity features based on statistical distribution disparities between ships and surrounding sea surfaces.  Mercier\textit{ et al.} \cite{mercier2006partially} modeled the wavelet coefficients of normal sea conditions as a zero-mean Gaussian mixture, combining wavelet-domain features with kernel functions for oil spill detection amid small-scale slicks and strong sea clutter. These methods remain severely constrained by background distribution priors and parameter estimation accuracy. Moreover, most are designed specifically for sea-surface ships, requiring domain adaptation or re-engineering when migrating to complex terrestrial contexts.

\textit{(iv) Discussion:} Despite the aforementioned limitations, as a classic framework for SAR target detection, CFAR has continued to evolve through integration with emerging technical paradigms such as deep learning \cite{lin2019dlcfar,li2023CFARADMM,faerch2025MSCFAR}. For instance, CFARnet \cite{diskin2024cfarnet} embeds CFAR constraints into the neural network architecture, enabling the model to learn a detector that complies with CFAR principles from data. CFAR-DP-FW \cite{zeng2024cfarDPFW} converts CFAR decisions into attention maps that are concatenated with the input of a convolutional network, enabling end-to-end training with a semantic loss. Other studies have applied CFAR to detection preprocessing \cite{jia2024fastShipDet} or clutter noise modeling \cite{wang2024RCGFDCFAR} to improve the generalization performance of detection systems in complex scenarios. This trend signifies a deep integration of statistical and data-driven approaches, offering new insights for target detection in complex environments.

\textbf{\textit{2) Non-Statistical Feature-based Methods:}} Beyond statistical methods, researchers leverage visual saliency, complex-domain physical features, or shallow learners for detection to circumvent clutter distribution priors. Wang \textit{et al.} \cite{wang2017saliency} used Bayesian saliency maps to preserve complete structures of targets and strong clutter, then employed morphological saliency maps combined with vehicle size priors to suppress natural and man-made strong clutter. Souyris \textit{et al.} \cite{souyris2004complex} and Ouchi \textit{et al.} \cite{ouchi2004shipcomplex} exploited coherence time differences between targets and clutter by dividing single-look complex imagery into sub-apertures along azimuth, enhancing weak scatterers via unnormalized Hermitian inner product or multi-look cross-correlation. Filippidis \textit{et al.} \cite{SAR2000PAMIDet} employed feedforward neural texture blocks for coarse target/non-target classification, fusing texture confidence, background discrimination, and size priors with fuzzy rules for airport aircraft detection. These approaches achieve robust detection in unknown or non-uniform clutter with lower computational costs than deep networks. These methods demand manual parameter tuning and specialization for specific scenes, requiring adaptive mechanisms or cascading with statistical features for complex terrestrial environments.

\subsection{Deep Learning-based SAR Target Detection}

This section, focusing on deep learning-based SAR target detection tasks, summarizes existing methods classified into anchor-based and anchor-free categories on their detection frameworks. The methods are categorized based on their attributes and core innovations, concluding with the key concerns discussion.

\begin{figure}[t]
\centering
\includegraphics[width=0.48\textwidth]{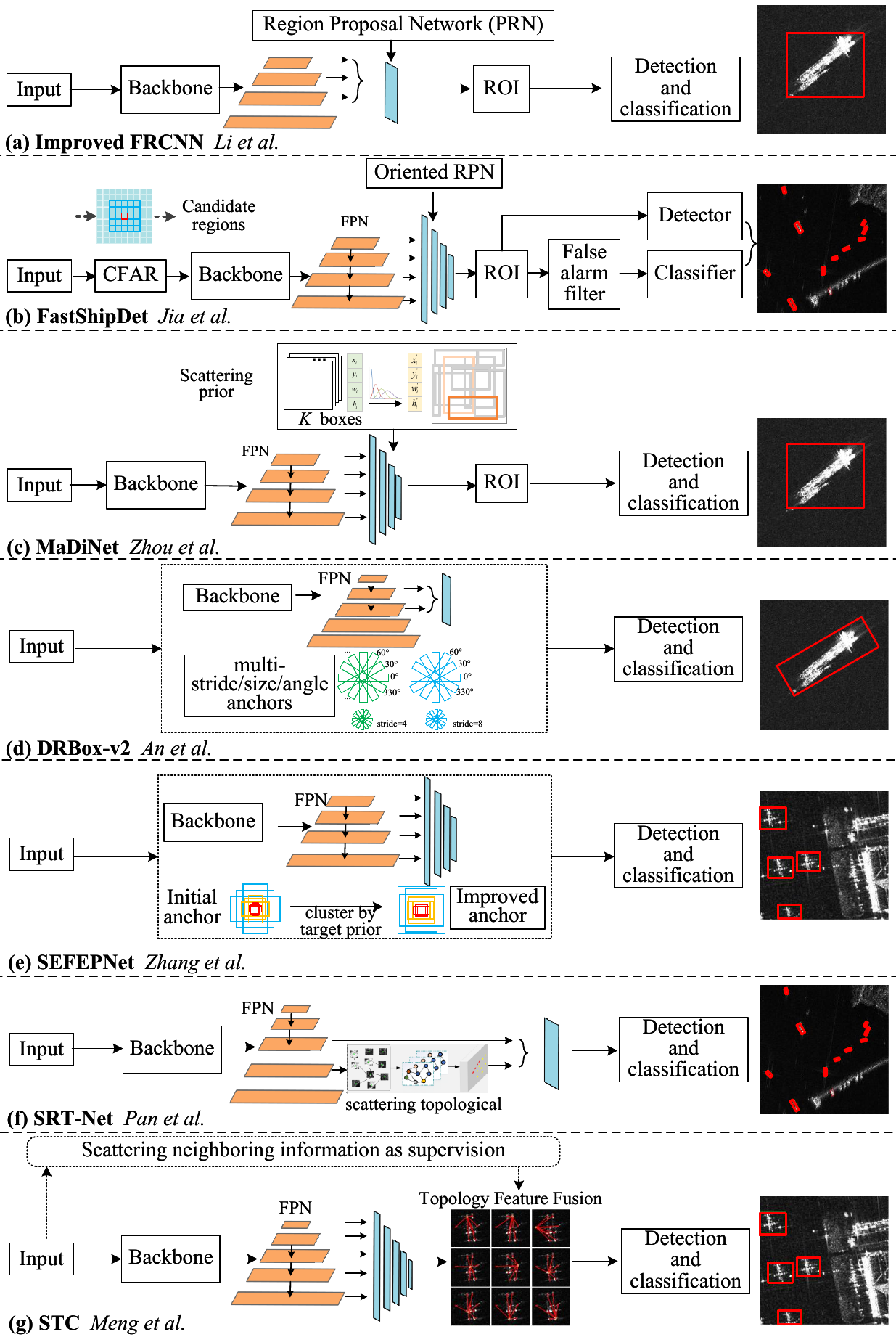}
\caption{Overview of representative deep learning-based methods for SAR target detection. (a)-(c) are anchor-based (two-stage) methods. (d) and (e) are anchor-based (one-stage) methods. (f) and (g) are anchor-free methods. ((a) Improved FRCNN \cite{li2017shipssdd}, (b) FastShipDet \cite{jia2024fastShipDet}, (c) MaDiNet \cite{zhou2025madinet}, (d) DRBox-v2 \cite{an2019drboxV2}, (e) SEFEPNet \cite{zhang2022sefepnet}, (f) SRT-Net \cite{pan2024srt}, (g) STC \cite{meng2025stcnet})}
\label{dl detection comparision}

\end{figure}

\begin{table*}[ht]
 \resizebox{0.9\linewidth}{!}{%
\begin{threeparttable} 
\setlength{\abovecaptionskip}{0.1cm} 
\setlength{\belowcaptionskip}{-0.1cm}
    \centering
         
    \caption{Performance of representative SAR target detection methods on six mainstream datasets. (SSDD \cite{li2017shipssdd}, SAR-Ship-Dataset \cite{wang2019sarshipdataset}, HRSID \cite{wei2020hrsid}, SAR-Aircraft-1.0 \cite{zhirui2023sarcraft}, SADD \cite{zhang2022saddsefepnet}, MSAR \cite{xia2022crtranssarMSAR}.)}
    \label{table:detection methods performance} 
\begin{tabular}{|c|c|c|c|c|c|c|c|c|c|c|} 
    \hline
        \multirow{2}*{Taxonomy} &  \multirow{2}*{Year} &  \multirow{2}*{Methods} & Open- & \multirow{2}*{Backbone} & \multicolumn{6}{c|}{Performance (mAP50,\%)} \\ 
        \cline{6-11} 
        & & & source& & SSDD & SAR-Shipdataset& HRSID & SAR-Aircraft1.0 & SADD  & MSAR \\ 
    \hline
        \multirow{12}*{anchor-based} & 2017 & Improved FRCNN \cite{li2017shipssdd}  &-& Z1F-Net & 78.8 & - & - & - & -  & - \\ 
    \cline{2-11}
        ~ & 2019 & DRBoxv2 \cite{an2019drboxV2} & \href{https://github.com/ZongxuPan/DrBox-v2-tensorflow}{Code} & VGG16 & 92.8 & - &- & - & - & - \\
     \cline{2-11}
        ~ & 2019 & YOLOv2-reduced \cite{chang2019yolov2reduced}&- & Darknet-19 & 90.0 & - & - & - & -  & - \\
  \cline{2-11}
        ~ & 2022 & SEFEPNet  \cite{zhang2022saddsefepnet} &\href{https://github.com/hust-rslab/SAR-aircraft-data}{Code} & Darknet-53 & - & - & - & - & \underline{\textbf{93.4}}  & - \\
    \cline{2-11}
        ~ & 2023 & HRLE-SARDet \cite{zhou2023hrle} &- & EfficientNet & 98.4 & - & 92.5  & - & - & \underline{88.4} \\
   \cline{2-11}
           ~ & 2024 &  ShipDetector \cite{li2024lightweightsar} &- & CSPNet & 97.6 & 91.2 & 93.6 & - & -  & - \\
    \cline{2-11}
           ~ & 2024 & DiffDet4SAR \cite{zhou2024diffdet4sar}  & \href{https://github.com/JoyeZLearning/DiffDet4SAR}{Code} & Res50+FPN & 96.9 & 95.1 & - & 88.4 & - & - \\
    \cline{2-11}
           ~ & 2024 & MSFA \cite{li2024sardet100} &\href{https://github.com/zcablii/SARDet_100K}{Code} &VAN & 97.9 & - & 83.7& - & -  & - \\
    \cline{2-11}
        ~ & 2025 & DAFDet \cite{yang2024dafdet} &- & Hybrid & 98.1 & \underline{96.5}  & - & - & - & \underline{\textbf{97.2}} \\ 
    \cline{2-11}
        ~ & 2025 & SARDet-CL \cite{yang2025sardetcl}&- & Res50 & - & - & -& 86.8 & \underline{87.7}  & 73.8 \\
     \cline{2-11}
    ~ & 2025 & PGD-YOLOv5 \cite{huang2025physics}& \href{https://github.com/XAI4SAR/PGD}{Code} & Res50 & - & -  & -  & 90.4 & -& - \\ \cline{2-11}
        ~ & 2025 & MaDiNet \cite{zhou2025madinet} & \href{https://github.com/JoyeZLearning/MaDiNet}{Code} & Hybrid & \underline{99.0} & \underline{\textbf{97.6}}  & -  & \underline{\textbf{90.8}} & -& - \\ 
    \hline
    \hline
        \multirow{13}*{anchor-free} & 2020 & SSE-ATD \cite{cui2020SSECenternet} & -&DLA & - & 94.7 & - & - & -  & - \\ 
    \cline{2-11}
        ~ & 2021 & FBR-Net \cite{Fu2020FERNet}  &-& Res50 & 94.1 & -& -  & - & - & - \\
 \cline{2-11}
        ~ & 2021 & CP-FCOS \cite{sun2021CPFCOS} &-& Res50 & - & - & \underline{96.0} & - & -  & - \\
     \cline{2-11}
        ~ & 2021 & Centernet++ \cite{guo2021centernet++} &-& DLA & 95.1 & 95.4& -  & - & - & - \\
   \cline{2-11}
        ~ & 2023 & SA-Net \cite{zhirui2023sarcraft}  &-& Res50 & - & - & - & 80.4 & - & - \\
    \cline{2-11}
        ~ & 2024 & 3SD-Net \cite{ma20243sdnet} &-& Res50 & 90.5 & 91.6 & - & - & - & - \\
   \cline{2-11}
        ~ & 2025 & PFARN \cite{li2025pfarn}&- & Res50 & 98.1 & - & 94.8  & - & - & - \\
   \cline{2-11}
        ~ & 2024 & MD-DETR \cite{liu2024MDDETR}&- & Swin-T+Res50 & 98.9 & -& -  & - & - & - \\
   \cline{2-11}
        ~ & 2024 & PVT \cite{chen2023pvt} &-& ViT & 96.8 & - & - & - & - & - \\
   \cline{2-11}
   ~ & 2024 & SFS-CNet \cite{li2024unleashing} & \href{https://github.com/like413/SFS-Conv}{Code}& CSPDarkNet &\underline{\textbf{99.6}} & - & \underline{\textbf{96.2}} & 89.7 & -  & - \\
   \cline{2-11}

        ~ & 2024 & STC-Net \cite{meng2025stcnet}  &- & Res50+FPN & - & -& - & 89.0 & -  & - \\
           \cline{2-11}
        ~ & 2025 & RDB-DINO \cite{qin2024rdbDINO}&- & DINO & 98.3 & - & 92.8 & - & - & - \\
        \cline{2-11}
      ~ & 2025 & PGD-YOLOv8 \cite{huang2025physics}& \href{https://github.com/XAI4SAR/PGD}{Code} & Res50 & - & - & - & \underline{90.7}  & - & - \\
    \hline
\end{tabular}

      \begin{tablenotes} 
		\item * The data are extracted from the original papers. We use ``-" to mark the dataset without reporting in the original papers. The \underline{\textbf{best}} results are \textbf{bold} and \underline{underlined}, while the \underline{second-best} are \underline{underlined} only. 
     \end{tablenotes} 
\end{threeparttable} 
}
     \vspace{-0.5cm}
\end{table*}

\textbf{\textit{1) Anchor-based SAR Target Detection:}} In SAR target detection, anchor-based methods facilitate localization and classification by presetting multiscale, aspect-ratio, and angle-varying bounding boxes. Based on their detection pipelines, they fall into two categories: two-stage and one-stage methods (Fig. \ref{detection comparision high level} (b) and (c)), which trade off detection accuracy against inference speed.

\textit{(i) Two-stage Methods:} Two-stage methods first generate candidate target regions via a Region Proposal Network (RPN), followed by fine-grained classification and location regression \cite{liu2020deep}. Improved Faster R-CNN \cite{li2017shipssdd} addresses the issues of multi-scale and dense distribution of ship targets by adopting multi-feature fusion to enhance target representation capability (Fig. \ref{dl detection comparision} (a)). SER Faster R-CNN \cite{lin2018SERFasterRCNN} incorporates the Squeeze-and-Excitation channel attention mechanism and a score correction strategy to improve the model’s ability to screen key scattering features. ARPN \cite{zhao2020arpn} utilizes a multi-branch convolutional structure to extract multi-scale features, aiming to tackle the problem of significant target size variations in SAR images. In recent years, emerging architectures such as transformers and diffusion models have also been integrated into the two-stage framework. Fast-ShipDet \cite{jia2024fastShipDet} applied the progressive detection process of global-regional-target to very large scenes (Fig. \ref{dl detection comparision} (b)). DiffDet4SAR \cite{zhou2024diffdet4sar} redefines the detection task as a bounding box denoising process, avoiding heuristic anchor box design. MaDiNet \cite{zhou2025madinet} builds on this by introducing a Gamma diffusion process to model the implicit association between the target position and scattering points and captures long-range dependencies utilizing the state-space model (Fig. \ref{dl detection comparision} (c)). This design improves the detection performance of structural targets in large scenes. These methods typically achieve high detection accuracy while incurring relatively high computational complexity.

\textit{(ii) One-stage Methods:} One-stage methods eliminate the region proposal step and directly perform target localization and classification simultaneously within the network \cite{wang2018SSDbasedATD,chang2019yolov2reduced}. As a result, they typically exhibit faster inference speed and are more suitable for real-time detection tasks. This category of methods achieves coverage of targets with varying sizes and orientations through dense anchor sampling and prediction across multiple feature levels. For instance, DRBoxv2 \cite{an2019drboxV2} proposes an improved rotated box encoding strategy and a multi-level prior box generation mechanism (Fig. \ref{dl detection comparision} (d)). It significantly enhances the detection accuracy for orientation-sensitive targets such as ships and aircraft. SEFEPNet \cite{zhang2022sefepnet}, on the other hand, redesigns anchor sizes based on prior knowledge of the scattering point distribution of aircraft targets, thereby improving the accuracy of target localization regression (Fig. \ref{dl detection comparision} (e)). Simultaneously, novel architectures continue to advance single-stage methodologies. MGCAN \cite{chen2022MGCAN} constructed a geospatial self-attention mechanism to enhance the modeling of contextual semantic relationships between targets and their surroundings. Additionally, lightweight designs are gaining traction. HRLE-SARDet \cite{zhou2023hrle} achieved high-precision multi-class detection with extremely low parameters. DAFDet \cite{yang2024dafdet} introduced a dynamic inference mechanism that adaptively adjusts computational paths based on image content, effectively balancing detection efficiency and accuracy.

\textit{(iii) Discussion:} Anchor-based SAR detection methods face three main challenges: limited generalization due to task-specific anchor designs, high computational overhead from dense anchor strategies, and mismatch imbalance in inhomogeneous scenarios. Future research should explore adaptive anchor mechanisms, lightweight designs, and the integration of physical knowledge to overcome these limitations.

\textbf{\textit{2) Anchor-free SAR Target Detection:}} Anchor-free methods eliminate the predefined anchor mechanism and achieve more flexible detection through keypoint detection, center point localization, or pixel-level prediction \cite{yu2023AFFDET,li2025pfarn} (Fig. \ref{detection comparision high level} (d)). SSE-CenterNet \cite{cui2020SSECenternet} integrates attention mechanisms in both channel and spatial dimensions to enhance semantic features. FBR-Net \cite{Fu2020FERNet} directly learns bounding box encoding to avoid the impact of anchor box bias. CP-FCOS \cite{sun2021CPFCOS} proposes generating guidance vectors from the classification branch to optimize the accuracy of localization regression. DenoDet \cite{dai2024denodet} integrates the transform-domain denoising concept from traditional image processing into the deep learning framework. By leveraging attention mechanisms to perform dynamic soft-thresholding in the frequency domain, it significantly enhances target detection accuracy and robustness in SAR imagery. To address the common issue of discontinuous target contours in SAR images, AFFDet \cite{yu2023AFFDET} adopts geometric projection to replace angle parameters. SRT-Net \cite{pan2024srt} extracts scattering points of aircraft targets via Harris corner detection and K-means clustering and constructs a graph structure to capture global information (Fig. \ref{dl detection comparision} (f)). SA-Net \cite{zhirui2023sarcraft} utilizes key scattering points for auxiliary localization, improving the detection reliability of aircraft targets. STC-Net \cite{meng2025stcnet} incorporates scattering topology cues into SAR aircraft detection and leverages their structural relationships to enhance robustness in complex scenarios (Fig. \ref{dl detection comparision} (g)). In recent years, DETR-based detection architectures have also made progress in the SAR field. For example, MD-DETR \cite{liu2024MDDETR} introduces a triple denoising strategy to achieve high-precision detection across multiple target categories. DET-Net \cite{suo2024DETNET} first unifies denoising, dynamic range compression, and channel combination into a single detection-based enhancement framework. RDB-DINO \cite{qin2024rdbDINO} explicitly constructs sample and noise queries during decoding, reducing Hungarian matching complexity and small target misdetection, thus enhancing matching efficiency and training stability. However, these anchor-free methods face inherent challenges, including stricter requirements for feature alignment and regression consistency, which complicate training and often lead to unstable convergence.

\textbf{\textit{3) Summary}}

\textit{(i) Performance Comparison:} This section systematically benchmarks mainstream SAR target detection methods. To ensure equitable comparison despite implementation variances (e.g., backbone architectures, feature fusion strategies, training protocols), we adopt mAP50 (\%) from six widely used public datasets (SSDD \cite{li2017shipssdd}, SAR-Ship-Dataset \cite{wang2019sarshipdataset}, HRSID \cite{wei2020hrsid}, SAR-Aircraft-1.0 \cite{zhirui2023sarcraft}, SADD \cite{zhang2022saddsefepnet}, MSAR \cite{xia2022crtranssarMSAR}) as the primary metric. To fully demonstrate the characteristics of each method, TABLE \ref{table:detection methods performance} provides their specific taxonomy and backbones. We have also provided codes of open-source methods for reproduction.

From the performance results, SFS-CNet \cite{li2024unleashing} achieved the best performance on the SSDD dataset with 99.6\%, followed closely by MaDiNet (99.0\%) \cite{zhou2025madinet}. On the SAR-Shipdataset, MaDiNet took the lead with 97.6\%, with DAFDet \cite{yang2024dafdet} trailing behind at 96.5\%. For the HRSID dataset, PFARN \cite{li2025pfarn} delivered excellent performance at 94.8\%, while CP-FCOS \cite{sun2021CPFCOS} also reached 96.0\%. On SAR-Aircraft1.0 and SADD datasets, SEEFNet \cite{zhang2022sefepnet} and PGD-YOLOv8 \cite{huang2025physics} achieved 93.4\% and 90.7\%, respectively, demonstrating their good generalization ability on specific target categories. As a multi-target scenario dataset, MSAR saw DAFDet \cite{yang2024dafdet} perform the best at 97.2\%, reflecting its outstanding cross-category detection capability.

\textit{(ii) Main Issues and Facts:} Evaluation frameworks often rely on singular metrics like mAP50, failing to adequately reflect overall performance across critical aspects such as missed/false detections, localization accuracy, and small target handling. Second, inconsistencies in experimental setups and implementation details (e.g., data augmentation, hyperparameters, and backbone) compromise comparison credibility. More importantly, the lack of publicly released code in most studies severely undermines reproducibility. Finally, a significant gap remains between current detection setups and real-world applications, as model generalizability under complex conditions, such as adverse weather, occlusion, and deformation, still lacks systematic verification.

%% file: Revision_noblue_section/EvolutionOfSARClassification.tex
\section{Evolution of SAR Target classification}
\label{sec:class}

\begin{figure}[t]
\centering
\includegraphics[width=0.49\textwidth]{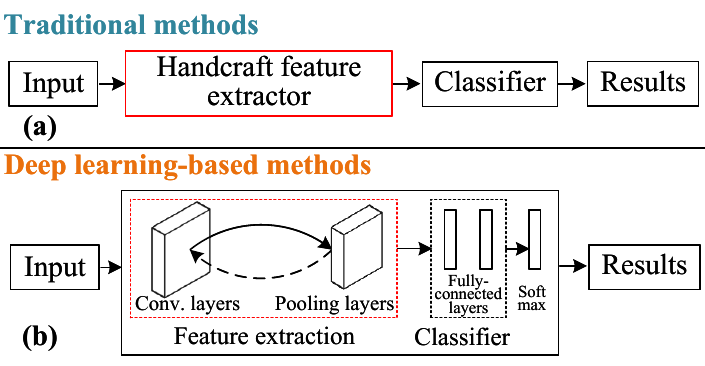}
\caption{Overview of key steps between traditional and deep learning-based methods for SAR target classification. (a) Traditional methods. (b) Deep learning-based methods.}
\label{classification comparision high level}
     \vspace{-0.5cm}
\end{figure}

\begin{figure}[ht]
\centering
\includegraphics[width=0.43\textwidth]{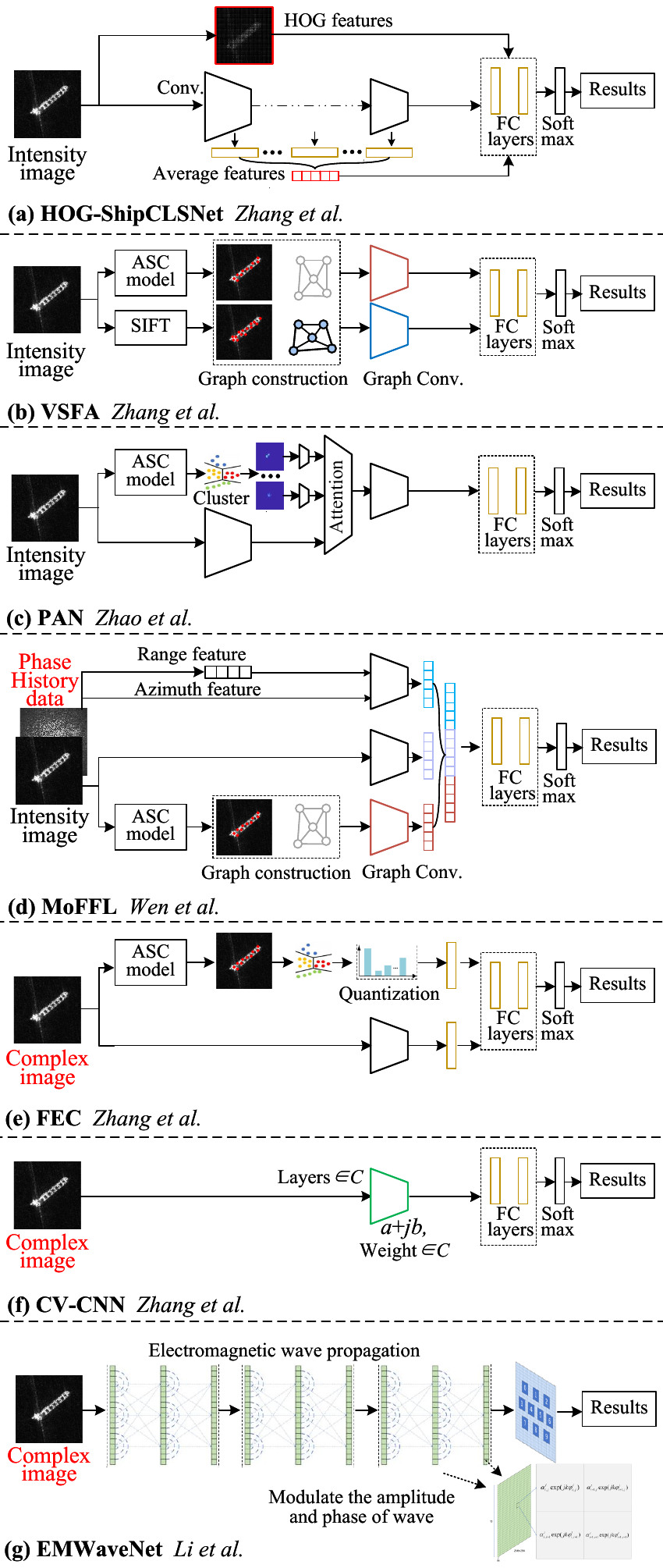}
\caption{Overview of representative deep learning-based methods for SAR target classification. (a)-(b) are intensity statistical feature-based, (c)-(d) are structural feature-based, and (e)-(g) are electromagnetic scattering feature-based methods. ((a) HOG-ShipCLSNet \cite{zhang2021hogShip}. (b) VSFA \cite{zhang2023vsfa}. (c) PAN \cite{feng2023pan}. (d) MoFFL \cite{wen2024mofel}. (e) FEC \cite{zhang2020fec}. (f) CV-CNN \cite{zhang2017complexCVCNN}. (g) EMWaveNet \cite{li2025emwavenet}.)}
\label{dl classification comparision}
     \vspace{-0.5cm}
\end{figure}

\textbf{The core of SAR target classification lies in extracting discriminative, effective, and robust features.} As shown in Fig. \ref{classification comparision high level} (a), traditional methods rely on handcrafted features combined with shallow classifiers. In contrast, deep learning methods automatically optimize feature representation via end-to-end learning (Fig. \ref{classification comparision high level} (b)). Despite paradigm differences, both aim to extract discriminative target features. From a feature representation perspective, this section presents a unified taxonomy and review of these methods, with critical issues discussed in each subsection.

\subsection{Traditional Methods}
We summarizes existing methods classified into intensity-based, texture-based, scattering modeling-based, and structural-based categories, noting some overlap across these domains. The methods are categorized based on the main feature utilized.

\textbf{\textit{1) Intensity-based Methods:}} These methods directly take the pixel intensity values of target region images as the source of features and high computational efficiency, such as PLF-SAR \cite{park2012PLF} and SPV-SAR \cite{chen2014spv}. Their fundamental assumption holds that targets with different structures and materials exhibit unique and stable backscattering statistical patterns under different azimuth angles (e.g., HMM-SAR \cite{papson2012clsShadow}). A typical practice involves extracting statistical metrics, such as mean, variance, histogram distribution, or moment features, from image slices, which are then input into traditional classifiers (e.g., SVM \cite{zhao2002SVMATR}, AdaBoost \cite{sun2007adaboost} or SAR-HOG \cite{song2016sarHOG}) for classification. For example, Enderli \textit{et al.} \cite{SAR2007PAMIRadar} proposed a SAR target classification method based on the weighted deflection criterion. By computing third-order pseudo-Zernike moments, the method used a quadratic filter to approximate the optimal likelihood ratio classifier. However, these methods are sensitive to noise and pose variations, often overlooking phase and contextual information. Consequently, they exhibit limited discriminative power and are prone to confusion among different targets under specific viewing angles.

\textbf{\textit{2) Texture-based Methods:}} Texture features leverage the spatial distribution patterns of pixel intensity in SAR images, with typical examples including the Gray-Level Co-occurrence Matrix (GLCM) and its derived metrics such as contrast, entropy, and correlation \cite{Cross1983texture,ulaby1986textural,soh2002SARtexture}. These features are used to characterize the roughness and uniformity of targets. Specifically, GLCM4Ice \cite{holmes1998textural} has been successfully applied to sea ice type discrimination, while MRF-SAR \cite{clausi2004comparingGLCPMRF,ochilov2012operationalSeaIce} systematically compared different texture modeling methods and analyzed the impact of window size on feature stability. However, the performance of texture-based methods is severely affected by speckle noise and exhibits poor robustness to target pose variations, which limits their application in complex scenarios.

\textbf{\textit{3) Scattering Modeling-based Methods:}} Based on the physical mechanism of electromagnetic scattering, this category of methods models target responses as a set of parameterized scattering centers, such as the Attributed Scattering Center (ASC) model based on the Geometric Theory of Diffraction (GTD) \cite{potter1997attributed} and TLS-SCR \cite{shen2006TLSESPRITSCR}. By fitting and extracting attributes including scattering point type, frequency, and azimuth-dependent factors, these methods form physically interpretable feature vectors, thereby elevating SAR image interpretation from the pixel level to the physical structure level. For such approaches, MSTAR-EOC \cite{keydel1996mstar} and GSC \cite{jianxiong2011MorphologicalOperation} systematically established evaluation criteria and a global scattering center model, respectively. This category of methods has laid a physical foundation for SAR target recognition and provided semantic priors for subsequent physics-guided deep learning. However, it suffers from limited ability to describe complex targets, high complexity of template matching, and a high degree of dependence on data quality.

\textbf{\textit{4) Structural Modeling-based Methods:}} Structural features focus on describing the macroscopic morphology and local invariant structures of targets. For instance, they can involve extracting contour regions through morphological operations \cite{jianxiong2011MorphologicalOperation}, or adopting SIFT descriptors \cite{dellinger2014sarSIFT,zHENG2018SIFT} from the optical field to extract rotation and scale-invariant features. NGCSE-ATR \cite{huang2014neighborhoodATR} proposes a feature representation method based on neighborhood geometric centers, which improves the performance of sample clustering. However, structural features in SAR images are susceptible to noise interference and relatively sensitive to local deformations and occlusions.

\textbf{\textit{5) Discussion:}} Despite these advances, traditional methods remain constrained by inherent limitations of handcrafted feature engineering. They heavily depend on expert prior knowledge, restricting generalization capabilities. Handcrafted features also suffer significant information loss, impairing their ability to characterize intra-class variations or complex target structures. Furthermore, the modular separation of detection, feature extraction, and classification prevents end-to-end collaborative optimization. These bottlenecks become especially prominent in complex scenarios, ultimately driving the shift toward data-driven, end-to-end deep learning solutions.

\subsection{Deep Representation Learning For SAR classification} 

Existing methods can be categorized into three categories according to the main types of information features used: intensity statistical feature-based, structural feature-based, and electromagnetic scattering feature-based methods.

\textbf{\textit{1) Intensity Statistical Feature-based Methods:}} These methods mainly extract apparent statistical and deep texture features from SAR intensity images. HOG-ShipCLSNet \cite{zhang2021hogShip} fused traditional HOG features with deep features extracted by convolutional networks (Fig. \ref{dl classification comparision} (a)). VSFA \cite{zhang2023vsfa} modeled the ASC and SIFT key points into graph structures (Fig. \ref{dl classification comparision} (b)) to fuse local scattering and spatial structure information. SAR-JEPA \cite{li2024sarjepa} replaced pixel reconstruction with gradient prediction and effectively overcame the interference of speckle noise on self-supervised learning. MIGA-Net \cite{wang2024MIGANet} used multi-view intensity images to model azimuth information in SAR sequences and improved angular robustness. Moreover, MJT-Net \cite{lv2024MJTNet} leverages the multi-head attention to mitigate view-induced feature inconsistency. Focusing on intensity-level feature enhancement and view modeling, these methods demonstrate strong applicability to complex structures and variable imaging angles, with robust engineering adaptability.

\textbf{\textit{2) Structural Feature-based Methods:}} These methods focuses on modeling spatial topological relationships within targets or between target components. It is particularly suitable for targets with explicit structural characteristics, such as vehicles and aircraft.  PAN \cite{feng2023pan} clustered ASC and introduced attention mechanisms, achieving component-level semantic and structural correlation (Fig. \ref{dl classification comparision} (c)). MoFFL \cite{wen2024mofel} proposed a hierarchical graph aggregation mechanism to construct target structural features from individual components to the entire target (Fig. \ref{dl classification comparision} (d)). LDSF \cite{xiong2025LDSF} introduced graph topological loss to enhance intra-class aggregation capability. MTSGL \cite{QishanHe2025MTSGLMS} incorporated structural templates and geometric transformations in aircraft classification, reducing reliance on pixel-level annotations. These methods explicitly utilize the spatial layout of targets and enhance robustness against structural variations and occlusions.

\textbf{\textit{3) Electromagnetic Scattering Feature-based Methods:}} These methods deeply explores the electromagnetic physical essence of SAR data, and can be further divided into complex domain modeling and physical mechanism embedding.

\textit{(i) Physical Mechanism Embedding:} These methods bridge interpretability and data-driven capabilities. FEC \cite{zhang2020fec} quantized ASC features and fused them with CNN deep features (Fig. \ref{dl classification comparision} (e)), enhancing target representation. CA-MCNN \cite{li2021CAMCNN}  integrates the ASC model into multi-scale CNNs, boosting robustness against occlusion and limited samples. PIHA \cite{huang2022piha} leverages high-level physical semantics to guide local feature learning. Recent advances like EMI-Net \cite{liao2024eminet} and ASC-U2Det \cite{wang2025ASCU2Det} further incorporate physical knowledge as supervisory signals, enforcing electromagnetic consistency across detection and classification tasks. This subcategory excels in operational scenarios demanding reliability and generalization, establishing a critical pathway for SAR target classification.

\textit{(ii) Complex Domain Modeling:} These methods construct complex neural networks to explore the complex characteristics of SAR data, thus improving the discrimination and robustness of target classification. CV-CNN \cite{zhang2017complexCVCNN} first proposed the complex-valued convolutional neural network to process both amplitude and phase information simultaneously (Fig. \ref{dl classification comparision} (f)). Subsequently, CV-FCNN \cite{yu2019cvfcnn} and MSCVNets \cite{zeng2022MSCVNets} further expanded the complex-valued convolutional structure by introducing fully convolutional and multi-scale mechanisms. In recent years, CV-SAR-Det \cite{wang2024cvsardet} proposed complex-valued loss functions and data augmentation strategies. FDC-TA-DSN \cite{gao2025fdc} designed four-dimensional dynamic weights to improve anti-noise performance. EMWaveNet \cite{li2025emwavenet} constructs an interpretable complex-valued network completely based on physical propagation formulas to promote the development of complex-valued networks toward physical interpretability (Fig. \ref{dl classification comparision} (g)). CRMC-Net \cite{2025CRMC-Net} and DAF-Net \cite{meng2025daf} optimized the complex-valued network structure from activation functions and view fusion, respectively. These methods optimize approaches by adapting to distinct target characteristics, emphasizing the intrinsic properties of SAR data at the signal level. It excels in tasks requiring sensitivity to electromagnetic attributes, offering robust theoretical foundations and strong framework extensibility.

\begin{table}[t]
\begin{threeparttable} 
    \setlength{\abovecaptionskip}{0.1cm} 
    \centering
    \caption{Performance of representative SAR target classification methods on MSTAR SOC \cite{keydel1996mstar}.}
      \vspace{-0.1cm}
    \label{table:CLASSIFICATION methods performance} 
    \begin{tabular}{|c|c|c|c|c|}
    \hline
  \multirow{2}*{Taxonomy}    & \multirow{2}*{Year}  & \multirow{2}*{Methods}   & Open-  & Performance \\ \cline{5-5}
              & &   & source  & (Acc,\%)  \\ \hline
 \multirow{6}*{Tradition}   &   2001  & SVM \cite{zhao2002SVMATR}  & -  & 90.00  \\ \cline{2-5}
       &  2001  & CondGauss \cite{o2002sarCondGaus}  &  - & 97.18  \\ \cline{2-5}
       &  2007  & AdaBoost \cite{sun2007adaboost}  & -  & 92.00  \\ \cline{2-5}
       &  2014  & DGM-ATR \cite{srinivas2014sarDGM}  &  - & 95.00 \\ \cline{2-5}
       &  2014  & MSRC \cite{dong2014MSRC}  &  - & 93.60  \\ \cline{2-5}
       &  2015  & MSS \cite{dong2015MSS}  & -  & 96.60  \\ \hline
     &  2016  & A-ConvNets \cite{chen2016target}  &  \href{https://github.com/jangsoopark/AConvNet-pytorch}{Code} & 99.13  \\ \cline{2-5}
      &   2017  & VDCNN \cite{pei2017VDCNN} & -  & 98.52  \\ \cline{2-5}
     Deep &   2020  & FEC \cite{zhang2020fec}  &  - & 99.59  \\ \cline{2-5}
    learning  &   2022  & SDFNet \cite{liu2022SDFNet} & -  & 99.58  \\ \cline{2-5}
       &  2023  & HDANet \cite{li2023HDANet} &  \href{https://github.com/waterdisappear/SAR-ATR-HDANet}{Code}  & \underline{99.64}  \\ \cline{2-5}
       &  2025  & MMFF \cite{lv2025mmff} &  - & \textbf{\underline{99.95}}  \\ \hline
    \end{tabular}
       \begin{tablenotes} 
		\item * The data are extracted from the original papers. Given that most existing literature employed inconsistent experimental setups and evaluation metrics, we prioritized methods that use the same training and test sets for comparison. The \underline{\textbf{best}} results are \textbf{bold} and \underline{underlined}, while the \underline{second-best} are \underline{underlined} only. 
     \end{tablenotes} 
\end{threeparttable} 
     \vspace{-0.5cm}
\end{table}

\textbf{\textit{4) Performance and Summary:}} We systematically summarize and compare mainstream SAR target classification methods on the classic MSTAR SOC dataset (TABLE \ref{table:CLASSIFICATION methods performance}). Deep learning methods show distinct advantages, with MMFF \cite{lv2025mmff} achieving near-perfect accuracy. However, critical challenges remain: evaluation relies heavily on singular accuracy metrics, failing to assess generalization and stability. Inconsistent experimental setups and limited code availability hinder comparability, reproducibility, and progress. High reported accuracies often reflect dataset-specific optimization rather than practical performance. Future research should focus on constructing high-quality, multi-scenario datasets to advance SAR target recognition toward real-world applications.

%% file: Revision_noblue_section/RecentAdvances.tex
\section{Recent Advances in SAR ATR}
\label{sec:recent advances}
 
Over the past three years, SAR ATR has experienced remarkable progress, driven primarily by four key aspects: foundation models, limited data, generative data enhancement and domain adaptation.

\subsection{Foundation models}

Foundation models \cite{bodnar2025foundation,yu2025Metaearth,zheng2025Changen2,wang2025HyperSIGMA}, pre-trained on extensive data in a task-agnostic manner (generally through self-supervised learning), can be flexibly adapted to a wide range of downstream tasks. Current research for SAR foundation models can be categorized into general foundation model empowerment, cross-remote-sensing industry models, and SAR-specific vertical models. (i) The general approach directly adapts models pretrained on natural images to SAR tasks \cite{tian2025visionsar}. It excels in segmentation by inheriting strong visual representations but struggles in classification and detection due to the domain gap between natural and SAR imagery. These methods also lack targeted designs for SAR-specific challenges. (ii) Cross-remote-sensing models aim for general representations across multi-source data (e.g., optical, SAR). SkySense++ \cite{wu2025skysense}, pretrained on massive multimodal data, achieves state-of-the-art results across 12 downstream tasks. However, it underexplores SAR detection and relies primarily on 10-meter resolution, single C-band data, limiting coverage of high-resolution and multi-band SAR. (iii) SAR-specific models focus on targeted designs. SARViT \cite{ma2023sarvit} validates ViT and masked autoencoding for SAR interpretation field. SARATR-X \cite{li2025saratrx} suppresses speckle noise via two-stage self-supervision. CV-SAR \cite{wang2025cvsar} embeds polarization decomposition into pretraining, and SUMMIT \cite{du2025summit} enhances scattering understanding via multi-task learning. These models outperform industry models in detection with smaller scales and better cost-effectiveness. However, their generalization in classification and segmentation still lags behind due to limited data scale and diversity.

\textbf{\textit{Discussion:}} However, current research still faces several fundamental challenges: (i) Model architectures lack SAR-specific design and exhibit insufficient adaptation to complex-valued data and non-Gaussian noise. (ii) Most models exhibit shallow modal utilization by relying solely on amplitude images and rarely exploring the scattering mechanisms contained in complex phase information. (iii) The scale and diversity of pre-trained data remain bottlenecks. Vertical domain models are limited by the amount of data, and industry models have insufficient coverage of high-resolution, multi-band SAR. (iv) A unified evaluation benchmark is lacking, where existing research overly relies on traditional tasks such as classification, detection, and segmentation while lacking systematic evaluation of high-value capabilities such as change detection, 3D reconstruction, and adversarial robustness. The deeper limitation is that current models seek breakthroughs strictly isolated within the SAR visual modality, systematically failing to assimilate global semantic priors (e.g., geospatial context, visual-linguistic priors, and electromagnetic laws). This cognitive confinement inhibits their in-depth understanding of the physical world, fundamentally limiting their reasoning and generalization capabilities in open-world scenarios.

\subsection{Limited data}

Target representation learning in SAR ATR is fundamentally challenged by the scarcity and imbalance of high-quality data, a multifaceted issue that extends beyond a simple lack of samples. This overarching problem of limited data encompasses not only the restricted quantity of annotated imagery but also the constrained scope of target categories and the imbalanced distribution of those categories. Collectively, these factors lead to unstable feature representations, decision boundary bias, and difficulties in generalization, coalescing into limited quantity, limited classes, and limited distribution.

\textbf{\textit{1) Limited Quantity:}} The primary challenge in few-shot SAR ATR stems from the severe scarcity of annotated training samples, a consequence of the high cost and expertise required for data labeling. This paucity of data makes it exceptionally difficult for models to learn discriminative features that generalize well. The problem is further compounded by the inherent complexity of SAR imagery, where phenomena like speckle noise and target variations can easily lead to overfitting on the few available examples \cite{Awais2025shipres}. To overcome this fundamental data shortage, current research primarily explores two avenues: advanced optimization strategies and data generation techniques \cite{huang2024generativeSAR, xv2025fewshot,zhang2025phgan}. (i) On the optimization front, methods like Mada-SGD \cite{zeng2023madasgd} enhance adaptability by learning key meta-parameters within the training process itself. Other approaches focus on improving sample efficiency, such as DCBES \cite{li2023DCBES}, which selects more representative samples through density clustering, and MBEN-BC \cite{gao2024mbenbc}, which refines classification by leveraging multiple levels of feature relationships. (ii) Data generation methods aim to synthetically expand the training set. For instance, SAR-INR \cite{cheng2025sarinr} learns an implicit representation to generate novel, continuous-view images, thereby enriching the limited data available for training.

\textbf{\textit{2) Limited Classes:}} A critical limitation in SAR ATR research is the narrow scope of target categories covered by current datasets, which fails to reflect the ever-expanding diversity of real-world targets. This constraint necessitates that models move beyond closed-set classification to handle the dynamic nature of practical applications. Consequently, research is increasingly focused on two key directions: class-incremental learning and open-set recognition, both aimed at extending the model's capability beyond its initial training scope. (i) Class-Incremental Learning: This approach addresses the scenario where new target categories are introduced over time. The core challenge is to learn these new classes without catastrophic forgetting of previously learned knowledge. Recent advances tackle this by designing training strategies that mitigate feature confusion, such as hybrid distance metrics \cite{wang2023heien}, handling aspect angle variations \cite{chen2025acrm}, or leveraging scene priors to reduce distribution bias in replay samples \cite{li2025CONTEXTER}. (ii) Open-Set Recognition: This direction confronts the reality that a model will inevitably encounter targets from unknown categories during deployment. The goal is to effectively distinguish known classes from unknown ones. Current methods focus on constructing discriminative spaces for unknown classes, for instance, via a reciprocity point mechanism \cite{xiao2025ASCRPL}, or on quantifying prediction uncertainty by estimating latent space likelihood using invertible flow models and GANs to identify unknown samples \cite{lin2025glge, gao2025DPSPGO}.

\textbf{\textit{3) Limited Distribution:}} Real SAR datasets are typically long-tailed: common targets substantially outnumber rare ones, biasing decision boundaries toward majority classes \cite{Awais2025samSAR,liu2025supervised}. To improve minority-class recognition, recent work has explored fairer losses and adaptive sampling, with meta-learning and curriculum learning emerging as promising paradigms. Meta-learning improves few-shot adaptation by optimizing meta-training, incorporating semantic guidance, or embedding physical knowledge. DSG \cite{zhou2025DSG} reduces base-class bias through semantic guidance and adversarial alignment, whereas the recent PGM \cite{li2026PGM} integrates amplitude–phase information and physics-based self-supervised pretraining to inject scattering priors and suppress few-shot speckle interference. Curriculum learning mitigates imbalance by adaptively ordering samples and weighting losses. Examples include easy-to-hard task decomposition for scattering-feature inversion \cite{chen2024RLSAR}, class-level hierarchical curricula with balanced sampling and dynamic weighting for long-tailed ship classification \cite{Muhammad2025Shipclass}, and TRANSAR’s \cite{yasin2025TranSAR} adaptive sampler, which initially oversamples foreground targets before gradually transitioning to batch-balanced sampling. Future work should integrate electromagnetic physics throughout the pipeline, establish end-to-end meta-learning–curriculum-learning frameworks, exploit unlabeled data through self-supervised learning, and develop adaptive training strategies and SAR-specific imbalance evaluation protocols.

\subsection{Generative Data Enhancement}

Data scarcity represents one of the core bottlenecks constraining the development of intelligent SAR image interpretation. Beyond few-shot learning, incremental learning, and open-set recognition, data augmentation techniques based on Generative Artificial Intelligence (GAI) offer a fundamental solution pathway. The core philosophy is to automatically synthesize highly realistic and fully annotated SAR samples by learning the distribution patterns of limited real SAR data or underlying physical imaging mechanisms, thereby expanding the scale and diversity of training sets and enhancing the generalization capability of downstream models. To address these challenges, existing research has primarily progressed along three directions.

\textit{\textbf{1) Physics- and Geometry-Aware Prior Embedding:}} The core idea of this paradigm is to explicitly embed the physical processes of SAR imaging or geometric prior information of targets into generative models to ensure physical consistency of generated results. GeoDiff-SAR II \cite{zhang2026geodiff,wu2026geodiffii} constructs a Geometric-Electromagnetic Conditional Map (GECM) that encodes target pose skeletons and dominant scattering centers into a structured intermediate representation. During training, GECMs are extracted from real images, while during inference they are directly rendered from 3D CAD models, providing strict geometric constraints for the diffusion model and enabling zero-shot extrapolation in missing large-angle intervals. Similarly, $\phi$-GAN \cite{zhang2025phgan} unrolls the ideal point scattering center model into a differentiable neural network module, effectively suppressing physical hallucinations. From a frequency-domain perspective, DiffuSAR \cite{ying2025DiffuSAR} leverages the high correlation between high-frequency components of SAR images and target scattering details, significantly improving generation fidelity without introducing excessive parameters.

\textit{\textbf{2) Disentangled Semantic Representation:}} SAR image formation arises from the joint action of multiple independent factors (geometry, scattering, texture, noise, angle). The disentangled representation paradigm decomposes these factors into independent latent space dimensions, modelling them separately before recombining for generation. Along this line, HuiYanEarth-SAR \cite{liu2026huiyanearth} disentangles macroscopic scene structure from microscopic scattering mechanisms, using AlphaEarth \cite{brown2025alphaearth} geographic semantics as priors for scene generation and a generative model for scattering-texture learning, thus enabling global SAR image synthesis. Under limited-data conditions, SAR-DisentDM \cite{yang2026sardisentdm} further decomposes SAR imagery into factors such as target semantics, background clutter, and speckle noise, reducing data dependence while improving generation diversity and realism.

\textit{\textbf{3) Task Utility Orientation: }}Unlike studies that focus primarily on image quality itself, a line of research designs generation strategies with the direct objective of improving downstream task performance. SAGA \cite{SAGAwu2026task} introduces a quality-assessment agent that integrates users’ natural-language requirements, the current status of heterogeneous datasets, and an available library of generation skills to automatically perform dataset characteristic analysis, augmentation strategy planning and execution, and result quality evaluation. X-Fake \cite{huang2025xFAKE} departs from traditional image quality assessment (IQA) and instead proposes to use a Bayesian neural network to estimate the uncertainty of generated samples, which serves as a quantitative indicator of distribution inconsistency. Moreover, SatDiFuser \cite{jia2025Satdifusercan} systematically explores multi-timestep, multi-scale fusion strategies for diffusion features and demonstrates that these generative features can outperform mainstream discriminative foundation models on remote sensing semantic segmentation and scene classification. This finding offers new insights for representation learning in the SAR domain: pre-training a diffusion model on large-scale unlabeled SAR data and subsequently employing it as a feature extractor for target recognition or detection promises to alleviate labeled data scarcity to some extent.

\textit{\textbf{Discussion:}} Despite notable progress, generative AI for SAR still faces several key challenges, including inadequate evaluation protocols, insufficient end-to-end integration of physical priors, the difficulty of balancing attribute disentanglement with generation diversity, and the limited theoretical understanding of feature-space augmentation. Existing metrics often fail to reflect downstream task utility, while physically grounded generation remains constrained by external modeling assumptions and data burdens. In addition, achieving precise controllability without sacrificing diversity, as well as clarifying the mechanisms and applicability of feature-level augmentation, will be crucial directions for future research.

\subsection{Domain Adaptation}

Cross-domain SAR ATR addresses distribution shifts from variations in imaging parameters, sensors, or modalities while preserving discriminative feature mapping \cite{zhao2026CDML}. Unlike traditional methods that assume independent and identically distributed data, SAR image statistics are highly sensitive to sensor heterogeneity, the simulation-to-reality gap, and changes in resolution, viewing angle, and modality. These factors complicate global feature alignment, erode domain-specific information, and increase inter-class similarity.
Early works relied on discriminators or gradient reversal layers to force feature distribution alignment between the source and target domains \cite{pan2023ida, zou2023UDAD,he2023CMFT}. Subsequent researchers decomposed domain differences into interpretable sub-problems, such as scattering topology \cite{zhang2025mfja}, rotation angle \cite{huang2024PITM}, and sub-aperture decomposition \cite{dong2024optisar}, and achieved differential compensation through gated fusion \cite{zhang2025mfja} or dynamic convolution \cite{zhang2025DFDM}. In terms of single-domain generalization, SAFA-MAO \cite{he2025safamao} adopted multi-gradient descent optimization to endow the model with meta-adaptability to changes in imaging conditions, with its loss function explicitly balancing task performance and domain invariance. CDFS-SAR \cite{shi2025cdfs} leveraged pre-trained natural image models and measures foreground-background separation via Brownian Distance Covariance to achieve zero-shot SAR knowledge transfer.

%% file: Revision_noblue_section/Futurework.tex
\section{Conclusion and Outlooks}
\label{sec:future}

\begin{figure}[t]
\centering
\includegraphics[width=0.495\textwidth]{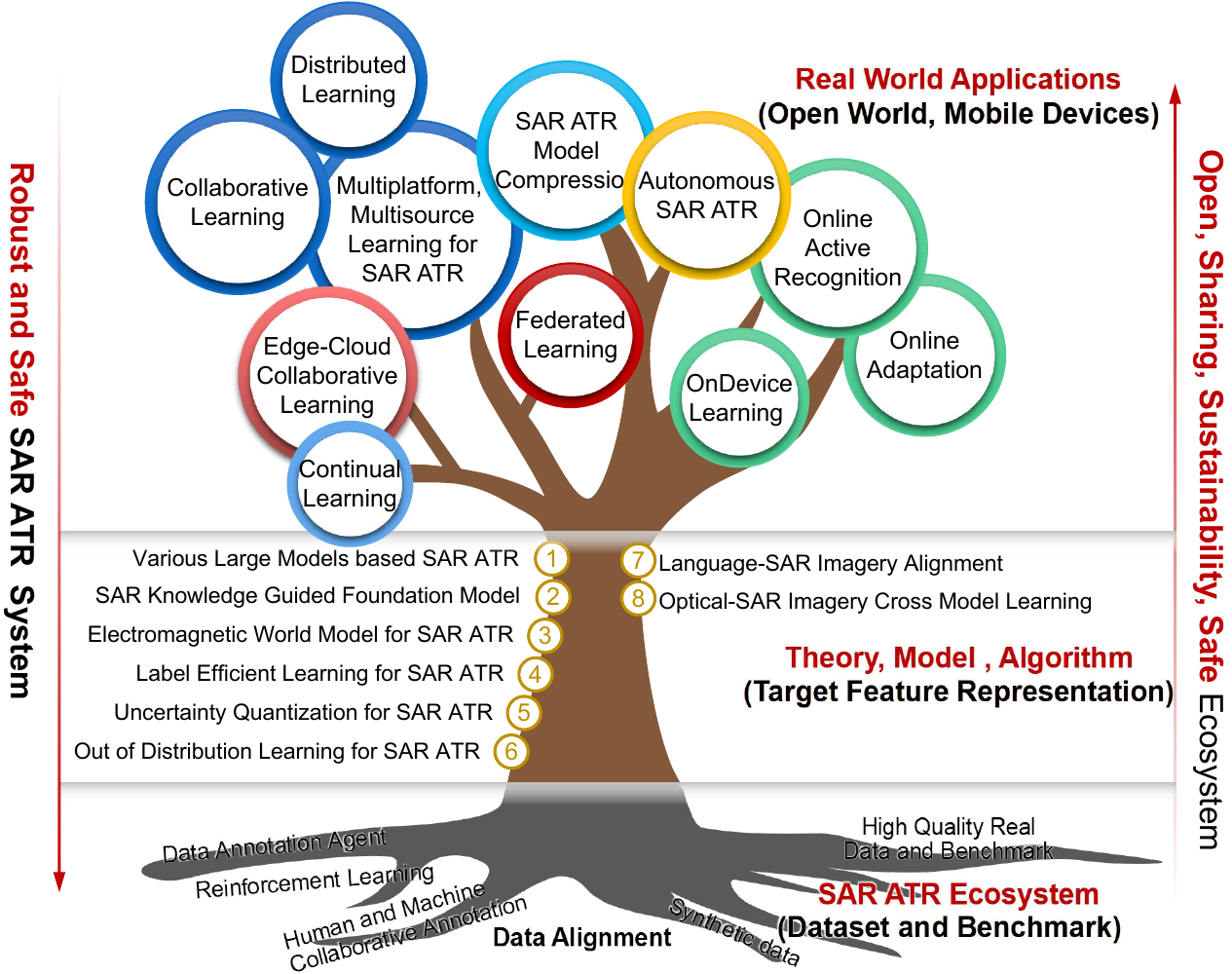}
  \vspace{-0.2cm}
\caption{Outlooks of SAR ATR across ecosystem; theory, model, and algorithm; real-world application; and system security.}
\label{futurework}
     \vspace{-0.5cm}
\end{figure}

After five decades of continuous evolution, SAR ATR stands at a critical juncture, transitioning from traditional informatization to advanced intelligentization and ultimately expanding toward autonomous systems. To unlock the full potential of AI in this domain, future research must shift toward a paradigm of bidirectional empowerment, not only applying AI to enhance SAR image understanding but also utilizing SAR physical mechanisms to constrain and guide AI models. As illustrated in Fig. \ref{futurework}, we propose a comprehensive and hierarchical technical roadmap visualized as a growing tree, encompassing four distinct dimensions: the foundational ecosystem, the core theory and model, the real-world applications, and the overarching system security.

\subsection{SAR ATR Ecosystem} 
Large-scale, standardized, high-quality data serves as the cornerstone for AI-powered SAR ATR. Unlike the domain of natural images, such as ImageNet, the SAR field has long faced challenges, including difficult data annotation and a scarcity of target-level data. While our team and others have conducted preliminary explorations \cite{liu2025sarmae,liu2025atrnet,zhu2026foundations}, we recommend that future breakthroughs be pursued in the followings.

\textbf{\textit{1) Intelligent Collaborative Annotation System:}} Given the acute shortage of trained SAR interpretation experts, an automated and efficient annotation paradigm is imperative. Future methodologies should incorporate data annotation agents \cite{swanson2025agentNature} driven by multimodal large models and reinforcement learning. These intelligent agents can autonomously execute preliminary screening and data cleaning, initiating an interactive human-machine collaborative annotation process that significantly improves label quality while reducing manual labor costs.

\textbf{\textit{2) Physics-Constrained Synthetic Data and Data Alignment:}} To address the long-tail distribution of measured data concerning extreme environments and rare targets \cite{chen2025opticalGe}, research should emphasize generating synthetic data driven by electromagnetic scattering mechanisms and SAR imaging processes (e.g., AlphaEarth \cite{brown2025alphaearth}). Concurrently, advanced data alignment techniques must be developed to minimize the domain shift between synthetic simulations and real-world observations, ensuring that models trained on synthetic datasets generalize effectively to actual measured data. This constructs an end-to-end high-fidelity simulation dataset covering scene-level, detection-level, and target-level tasks, transforming the field from data scarcity to a data-rich mine.

\textbf{\textit{3) High-Quality Real Data and Standardized Benchmark:}} The community must prioritize the collection and curation of high-quality real measured data to overcome current data limitations. Furthermore, establishing standardized, open-source benchmarks covering multiple complexity levels is essential \cite{center2026benchmark}. This ensures that algorithmic evaluations, ranging from accuracy to robustness, are conducted under unified test protocols, fostering equitable technological comparisons.

\subsection{Theory, Model and Algorithm}

Building upon a robust data ecosystem, the fundamental scientific challenge is achieving accurate, robust, and interpretable SAR target feature representation in complex, open environments. Future theories must evolve from pure data-driven mapping toward physics-interpretable architectures.

\textit{\textbf{1) Various Large Models based SAR ATR:}} Constrained by data and computing resources, SAR ATR should adopt a pragmatic development path from point to surface to volume. Initially, domain-specific foundation models \cite{li2025saratrx, yang2025fusarklip} can be established to build adaptation expertise. Subsequently, these models can be extended to industry-level general SAR foundation models, fusing multi-source SAR data for cross-scenario generalization. Ultimately, the focus shift toward multimodal large models that integrate optical, textual, and geospatial data.

\textit{\textbf{2) SAR Knowledge Guided Foundation Model:}} By integrating scattering topology and phase information, we create foundation models that are physically interpretable rather than black boxes. Specific approaches include mining sparse scattering information using pixel differencing and masking techniques, fusing statistical distribution priors of SAR data, and designing complex-valued neural networks and physical neural networks \cite{momeni2025trainingPNN} that conform to the characteristics of the complex domain. 

\textit{\textbf{3) Electromagnetic World Model for SAR ATR:}} Target recognition entails perceiving the objective world. Future efforts must move beyond imagery by building electromagnetic world models to achieve refined SAR target inversion. Furthermore, integrating embodied intelligence \cite{mon2025embodied} casts foundation models as brains and agents as hands, enabling continuous learning through interaction and the evolution of capabilities from static recognition to dynamic adaptation.

\textit{\textbf{4) Label Efficient Learning for SAR ATR:}} Label-efficient learning is the key to overcoming the scarcity of annotated data in SAR ATR. Through paradigms such as active learning, few-shot learning, zero-shot learning, and self-supervised pretraining, it maximizes the use of limited annotation resources, enabling models to generalize from minimal samples.

\textit{\textbf{5) Uncertainty Quantization for SAR ATR:}} Uncertainty quantification equips SAR ATR models to recognize their own limitations, transforming them from a blind state of mismatched confidence and accuracy into intelligent systems capable of perceiving prediction reliability. When facing out-of-distribution targets or severe noise interference, the model can issue warnings through high uncertainty rather than forcing incorrect outputs.

\textit{\textbf{6) Out of Distribution (OOD) Learning for SAR ATR:}} OOD learning equips SAR ATR models to delineate their cognitive boundaries, allowing them to determine if inputs fall outside the training distribution. Faced with unseen scenarios or anomalies, the model triggers alerts instead of forcing inference beyond its limits, thus preventing uncontrollable decision failures. This ensures that the recognition system maintains stable performance when encountering novel targets or unseen background clutter

\textit{\textbf{7) Language-SAR Imagery Alignment:}} By mapping SAR images and text into a unified semantic space, models can leverage language to introduce semantic priors, enabling zero/few-shot recognition with limited annotations. This alignment shifts the paradigm from static classification to interactive cognition, where users query via natural language and receive semantic descriptions instead of mere class labels, enhancing interpretability.

\textit{\textbf{8) Optical-SAR Imagery Cross-Model Learning:}} Cross-modal learning maps data into a unified representation space for complementary insights. Optical modalities offer semantic priors that help suppress SAR speckle noise and reconstruct missing structures, while SAR provides all-weather, day-and-night capabilities to enhance optical robustness. Crucially, this alignment enables knowledge transfer from large-scale labeled optical data to SAR, mitigating the high SAR annotation costs.

\subsection{Real-World Application}

Theoretical and algorithmic innovations must ultimately translate into practical capabilities for diverse deployment platforms. The application layer focuses on bridging theoretical models with multi-platform hardware constraints.

\textbf{\textit{1) Collaborative and Distributed Intelligence: }}
For distributed sensor networks, such as satellite constellations and autonomous  aerial vehicle swarms, architectures relying on centralized processing are no longer viable. Future systems must integrate distributed learning, collaborative learning, and multi-platform, multisource learning to fuse heterogeneous sensor data efficiently. Moreover, the implementation of federated learning will enable multiple distributed platforms to collaboratively train a global SAR ATR model while keeping raw data strictly localized, optimizing communication bandwidth and ensuring data privacy. This facilitates a seamless edge-cloud collaborative learning infrastructure \cite{bai2025decade}.

\textbf{\textit{2) Autonomous and Online Intelligence:}}  This enables SAR ATR systems to operate independently in dynamic, open-world environments. Deploying models on resource-constrained mobile devices requires model compression techniques, such as quantization and pruning, to facilitate efficient on-device learning and inference. Post-deployment, online adaptation capabilities allow continuous adjustment to changing environments. Complementing this, online active recognition strategically queries human experts for ambiguous targets, establishing an efficient human-in-the-loop. To ensure viability, continual learning prevents catastrophic forgetting, enabling models to acquire new knowledge while preserving old representations. Together, these capabilities pave the way for fully autonomous SAR ATR systems capable of sustained, adaptive operation without continuous human oversight.

\subsection{SAR ATR System Security}
As SAR ATR algorithms evolve from laboratory-controlled scenarios to mission-critical practical deployments, guaranteeing the overall trustworthiness of relevant systems has become an indispensable requirement instead of an auxiliary optimization, and security and trustworthiness need to be strictly implemented throughout the full-stack responsible system covering data, research methodologies and practical applications \cite{nature2025resp}. For SAR ATR datasets, robust defenses are essential to counter data injection during collection, prevent storage leaks, and mitigate generated data contamination \cite{shumailov2024ai}. For SAR ATR models, robustness should be enhanced via adversarial training, while integrated uncertainty quantification and interpretability mechanisms ensure decision traceability. For applications, the physical security of edge nodes, the authenticity and integrity of human-machine interaction, and trustworthy multi-platform collaboration must be strictly guaranteed. Ultimately, a comprehensive evaluation framework for trustworthiness must be established. This framework should cover data quality, model robustness, and system attack resistance to certify the system for large-scale deployment in critical missions.

%% file: Revision_noblue_section/bio.tex
\footnotesize

\vspace{-12mm}
\begin{IEEEbiography}[{\includegraphics[width=1in,height=1.25in,clip,keepaspectratio]{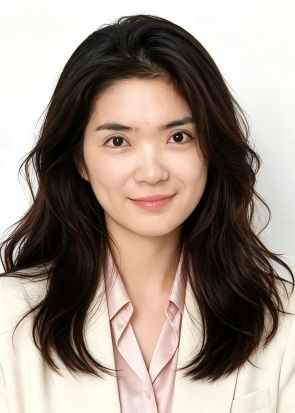}}]{Jie Zhou} received the B.S. degree from the National University of Defense Technology, Changsha, China, in 2020, where she is currently pursuing the Ph.D. degree with the College of Electronic Science and Technology. Her research interests include generative models, remote sensing image interpretation, and SAR target recognition.
\end{IEEEbiography}

\vspace{-15mm}
\begin{IEEEbiography}[{\includegraphics[width=1in,height=1.25in,clip,keepaspectratio]{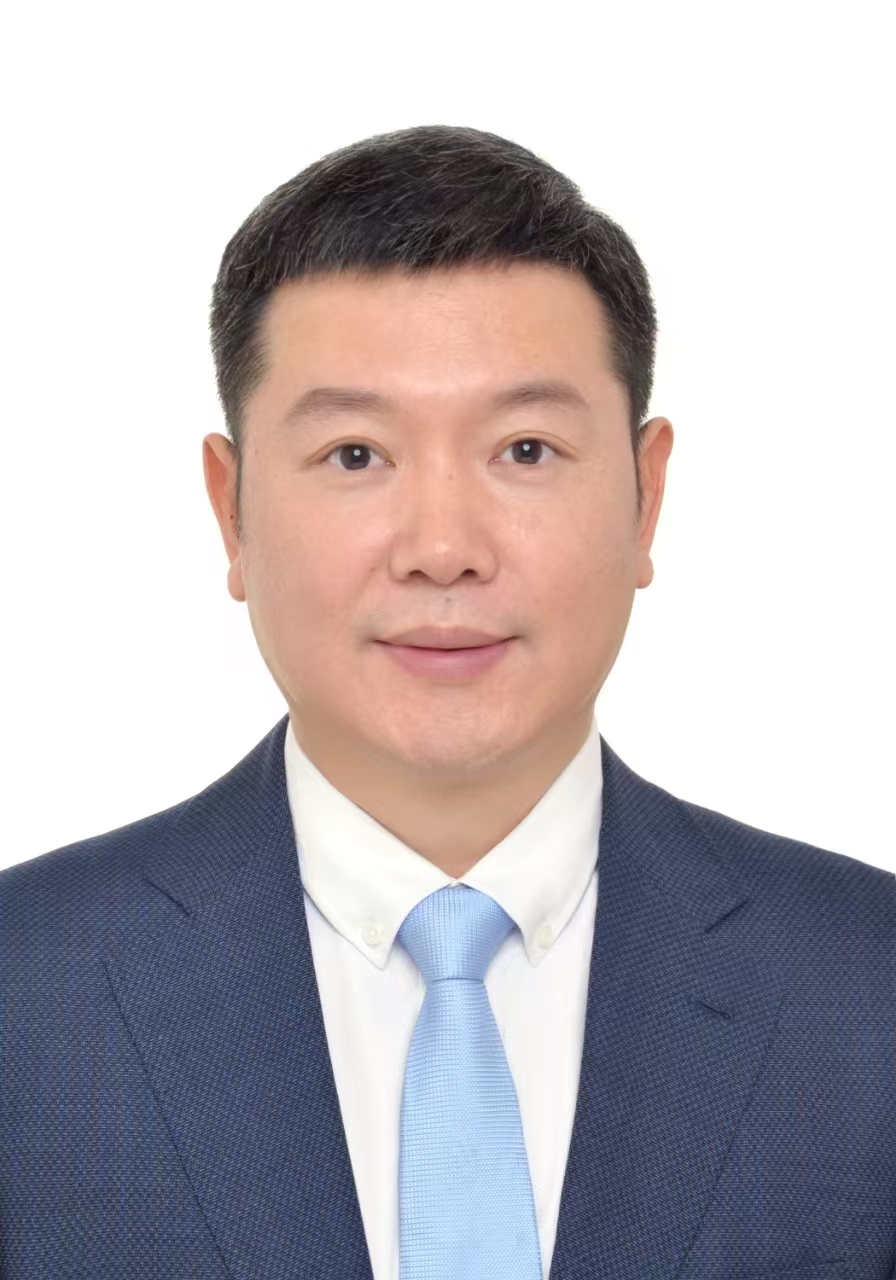}}]{Yongxiang Liu} (Member, IEEE) received the B.S. and Ph.D. degrees from the College of Electronic Science at the National University of Defense Technology, Changsha, China, in 1999 and 2004, respectively. He is currently a Full Professor at the National University of Defense Technology. He has published numerous papers in respected journals, including IEEE TPAMI, IEEE TIP, and IEEE TGRS. His research interests mainly include radar imaging, SAR image interpretation, and artificial intelligence.
\end{IEEEbiography}

\vspace{-15mm}
\begin{IEEEbiography}[{\includegraphics[width=1in,height=1.25in,clip,keepaspectratio]{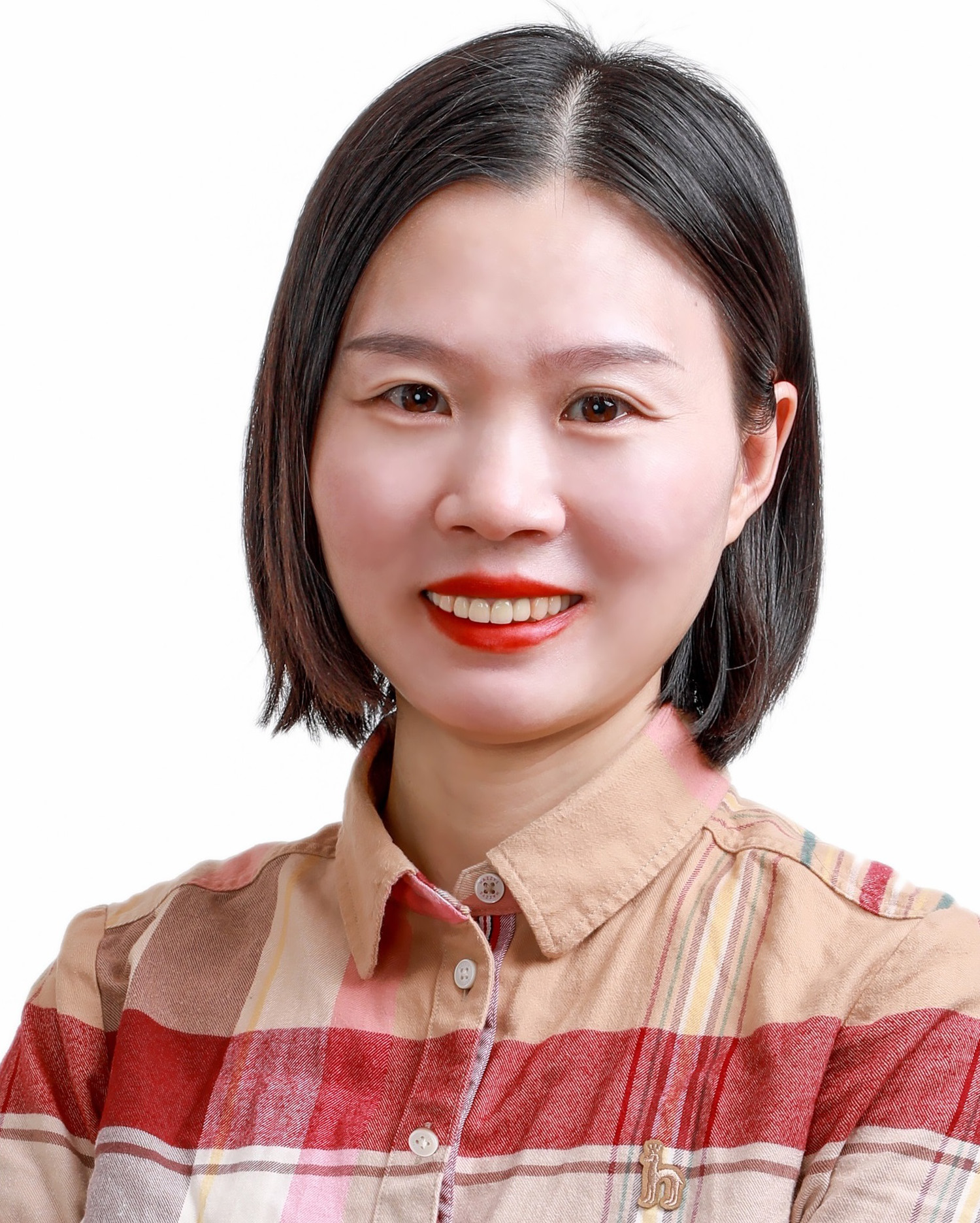}}]{Li Liu }  received her Ph.D. from National University of Defense Technology, China, in 2012 and is now a Full Professor there. She has visited the University of Waterloo, Chinese University of Hong Kong, and University of Oulu. She has co-chaired workshops for CVPR and ICCV, served as lead guest editor for IEEE TPAMI and IJCV, and is an Associate Editor for IEEE TCSVT and PR. Her research in computer vision, pattern recognition, and machine learning has garnered over 25,000 citations.
\end{IEEEbiography}


\vspace{-15mm}
\begin{IEEEbiography}[{\includegraphics[width=1in,height=1.25in,clip,keepaspectratio]{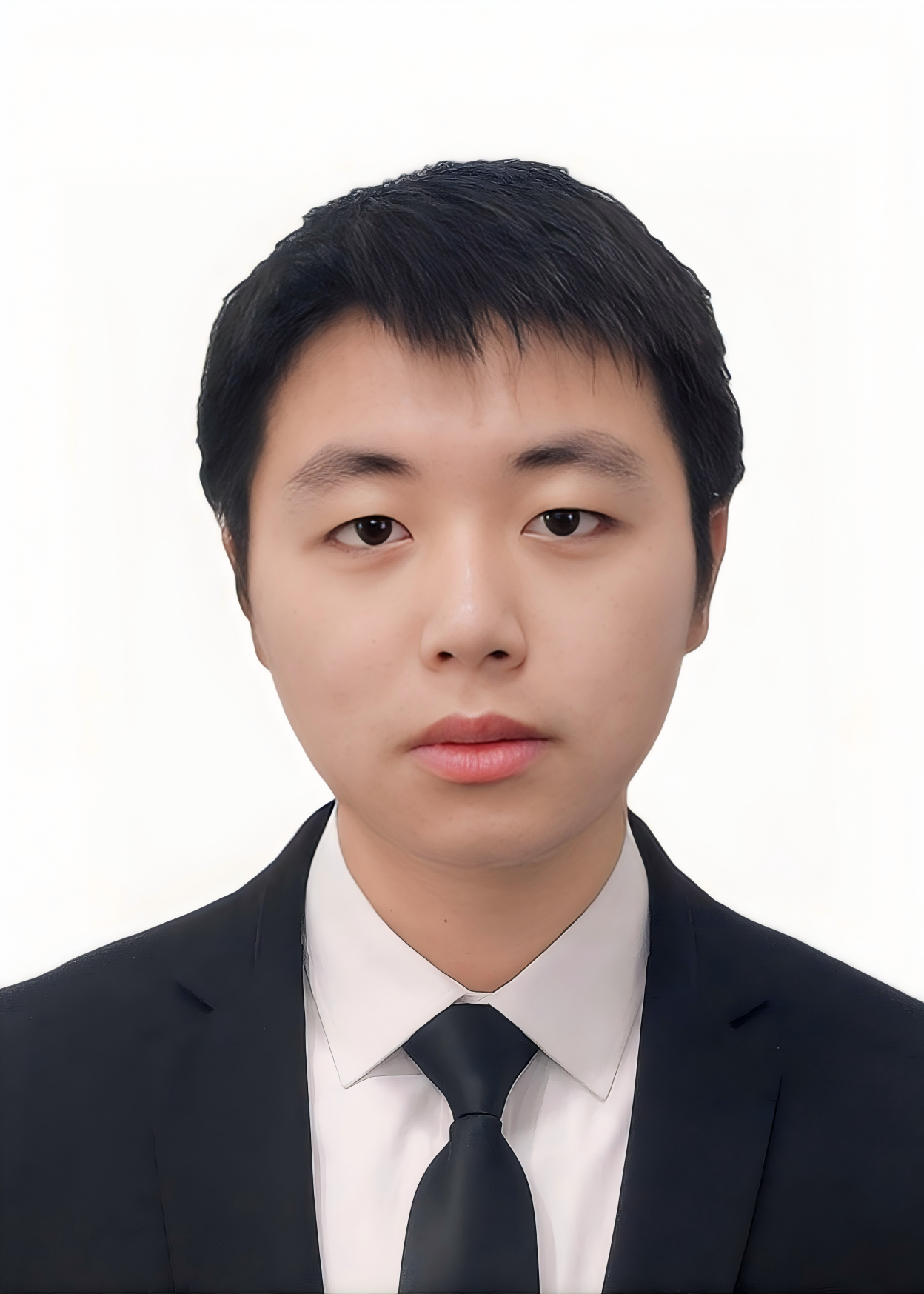}}]{Weijie Li} received his Ph.D. degree from the National University of Defense Technology, Changsha, China. He has published papers in respected journals, including IEEE TIP and ISPRS JPRS. His research interests mainly focused on radar target recognition and deep learning. 
\end{IEEEbiography}

\vspace{-15mm}
\begin{IEEEbiography}[{\includegraphics[width=1in,height=1.25in,clip,keepaspectratio]{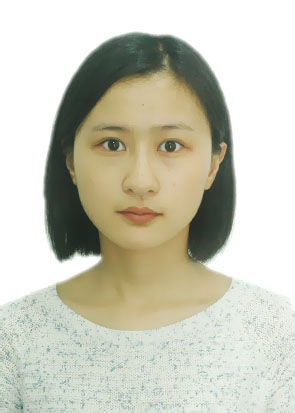}}]{Taoli Yang} (Senior Member, IEEE) received the Ph.D. degree in signal processing from the National Key Lab of Radar Signal Processing, Xidian University, Xi’an, China, in 2014.  From 2015 to 2016, she was a Postdoctoral Research Fellow with the School of Electrical and Electronic Engineering, Nanyang Technological University, Singapore. She is currently an Associate Professor with the School of Resources and Environment, University of Electronic Science and Technology of China, Chengdu, China. Her current research interests include SAR/ISAR imaging, interferometric SAR, GRACE, and ground moving target indication.
\end{IEEEbiography}

\vspace{-15mm}
\begin{IEEEbiography}[{\includegraphics[width=1in,height=1.25in, clip,keepaspectratio]{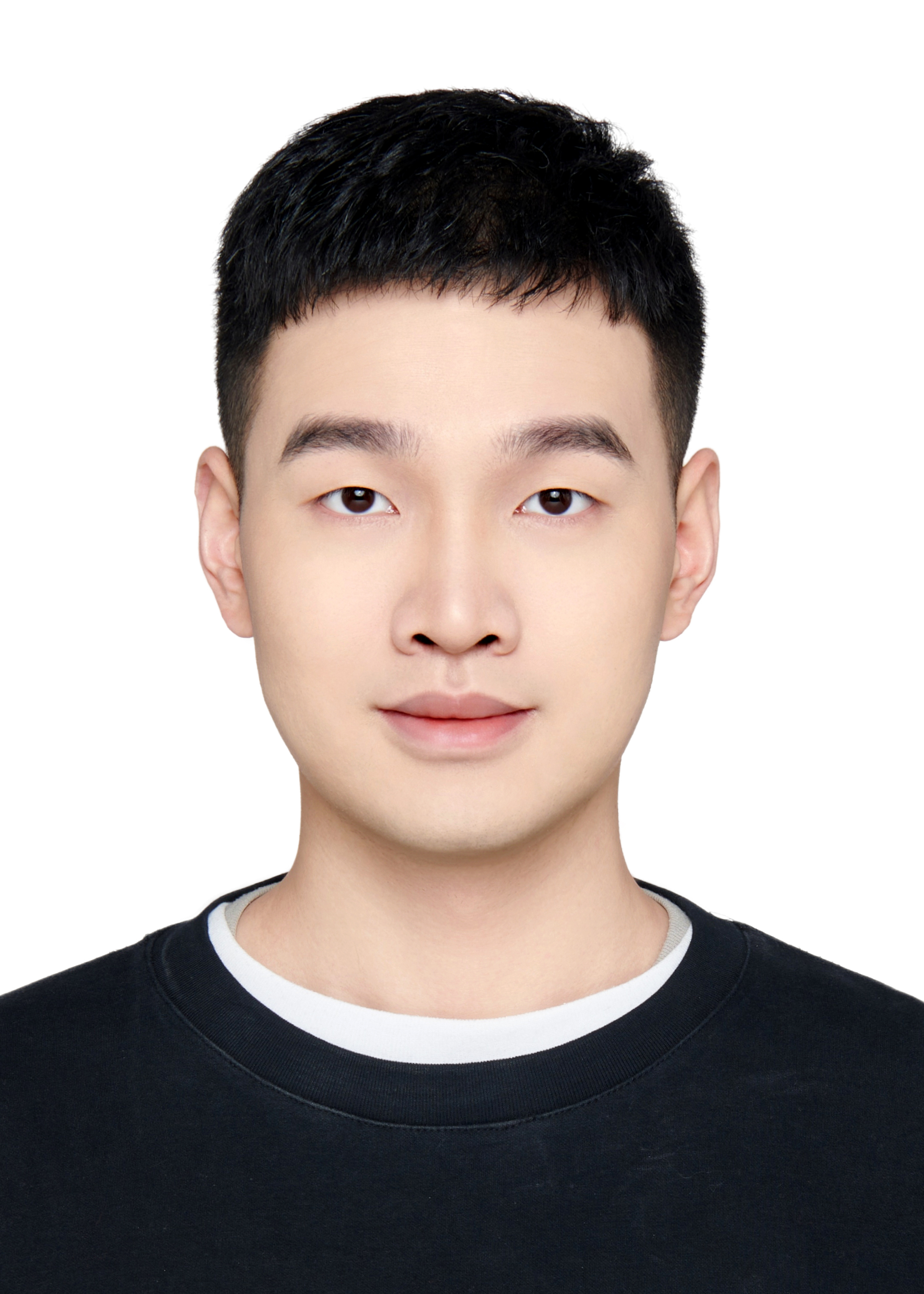}}]{Bowen Peng} received his B.S. and M.S. degrees from the National University of Defense Technology, Changsha, China, respectively in 2020 and 2022. He is pursuing the Ph.D. degree with the Comprehensive Situation Awareness Group, College of Electronic Science and Technology, National University of Defense Technology. His research interests include adversarial robustness of deep learning and trustworthy remote sensing object recognition.
\end{IEEEbiography}

\vspace{-15mm}
\begin{IEEEbiography}[{\includegraphics[width=1in,height=1.25in, clip,keepaspectratio]{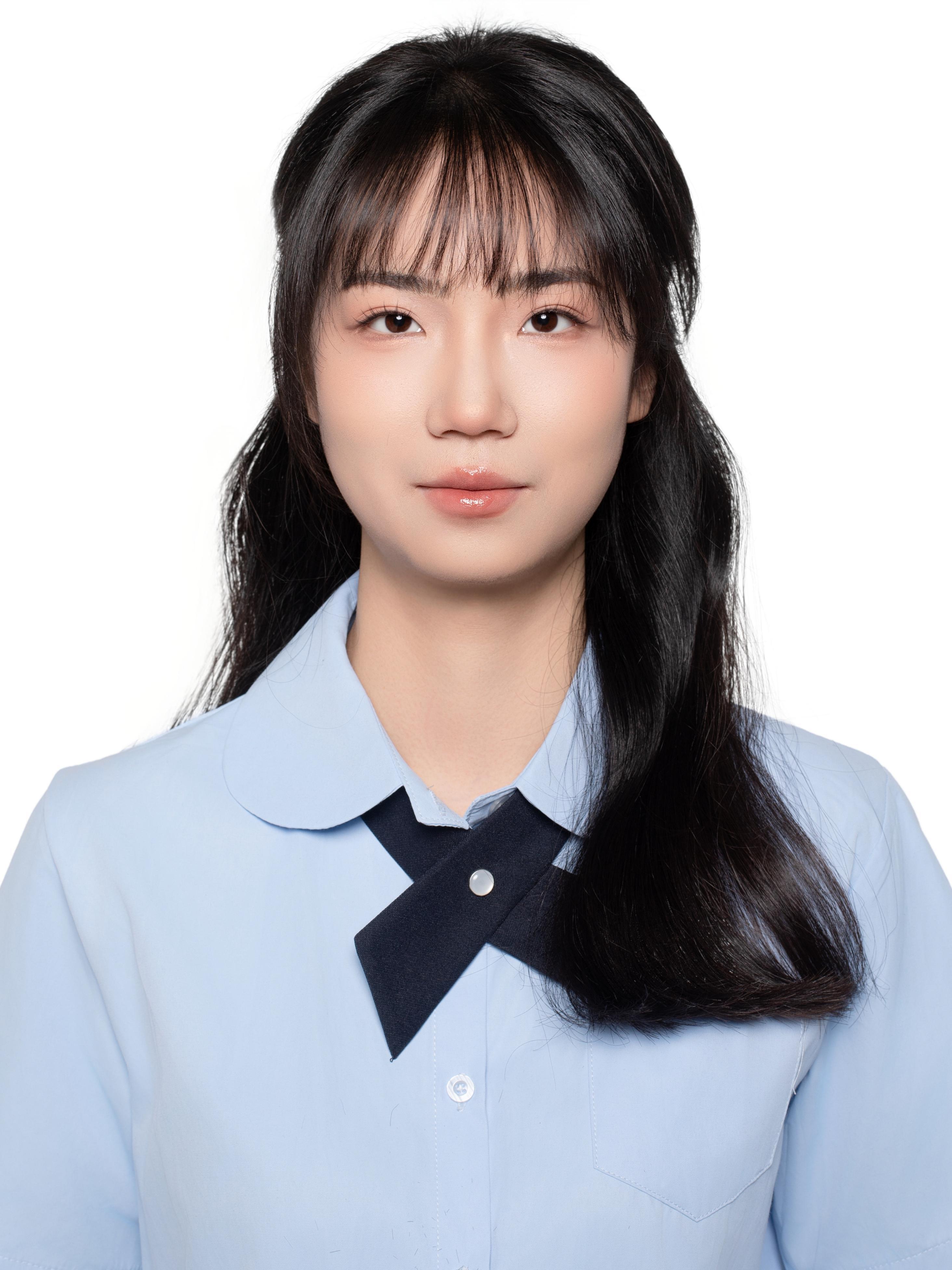}}]{Yafei Song} received the M.S. degree in the college of electronical and informatiom engineering from Xidian University, Xi'an, China, in 2024. She is currently pursuing the Ph.D. degree in information and communication engineering with the National University of Defense Technology, Changsha, China. Her main research interests include deep learning and its application in remote sensing image analysis.
\end{IEEEbiography}

\vspace{-15mm}
\begin{IEEEbiography}[{\includegraphics[width=1in,height=1.25in, clip,keepaspectratio]{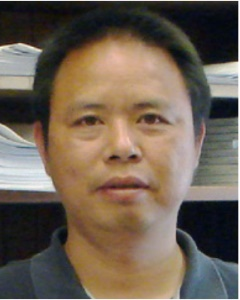}}]{Gangyao Kuang} received the B.S. and M.S. degrees in geophysics from the Central South University of Technology, Changsha, China, in 1988 and 1991, respectively, and the Ph.D. degree in communication and information from the National University of Defense Technology, Changsha, in 1995. He is currently a Professor with the College of Electronic Science, National University of Defense Technology. His research interests include remote sensing, SAR image processing, change detection, SAR ground moving target indication, and classification with polarimetric SAR images.
\end{IEEEbiography}

\vspace{-15mm}
\begin{IEEEbiography}[{\includegraphics[width=1in,height=1.25in,clip,keepaspectratio]{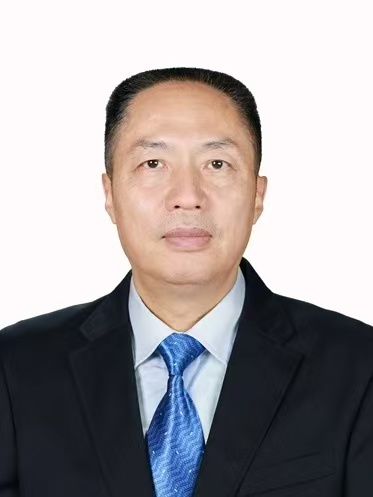}}]{Xiang Li} received the B.S. degree in electronic engineering from Xidian University, Xian, China, in 1989 and the Ph.D. degree in information and communication engineering from National University of Defense Technology, Changsha, China, in 1998. He is currently a Professor at the National University of Defense Technology. His research interests include signal processing, automation target recognition and deep learning.
\end{IEEEbiography}